\documentclass[aps,prd,onecolumn,superscriptaddress,nofootinbib,showpacs]{revtex4-1}

\usepackage{bm}
\usepackage{amsmath}
\usepackage{natbib}
\usepackage{graphicx}
\usepackage{comment}

\def\ba{{\bm a}}
\def\bb{{\bm b}}

\def\bk{{\bm k}}

\def\bn{{\bm n}}
\def\bN{{\bm N}}

\def\bv{{\bm v}}

\def\bx{{\bm x}}

\def\by{{\bm y}}
\def\bD{{\bm D}}
\def\bR{{\bm R}}

\def\memsai{Memorie della Societ\`a Astronomica Italiana} 
\def\planss{Planetary and Space Science}

\def\ba{{\bm a}}
\def\bb{{\bm b}}

\def\bk{{\bm k}}

\def\bn{{\bm n}}
\def\bv{{\bm v}}

\def\bx{{\bm x}}

\def\by{{\bm y}}
\def\bz{{\bm z}}
\def\bD{{\bm D}}
\def\bN{{\bm N}}
\def\bR{{\bm R}}

\def\lb{\label}
\def\be{\begin{equation}}
\def\ee{\end{equation}}
\def\bea{\begin{eqnarray}}
\def\eea{\end{eqnarray}}

\def\lb{\label}

\def\be{\begin{equation}}
\def\ee{\end{equation}}
\def\bea{\begin{eqnarray}}
\def\eea{\end{eqnarray}}

\newcommand{\gab}[3]{g^{#1 #2}_{(#3)}}

\newcommand \dDdx[1]{\frac{\partial{\Delta}^{(1)}_{r}}{\partial x^{#1}} (\bm z(\lambda),t_B,\bx_B)}
 
\newcommand \dDdxa[1]{\frac{\partial{\Delta}^{(1)}_{r}}{\partial x_{A}^{#1}} (\bm z(\lambda),t_B,\bx_B)} 
\newcommand \dDdxb[1]{\frac{\partial{\Delta}^{(1)}_{r}}{\partial x_{B}^{#1}} (\bm z(\lambda),t_B,\bx_B)} 

\newcommand{\dgab}[4]{g^{#1 #2}_{(#3),#4}}

\def\bx{{\bm x}}  
\def\ba{{\bm a}}  
\usepackage[colorlinks=true]{hyperref}   

\begin{document}

\title{Relativistic formulation of coordinate light time, Doppler and astrometric observables up to the second post-Minkowskian order }




\author{A. Hees}
\email{aurelien.hees@gmail.com}
\affiliation{Jet Propulsion Laboratory, California Institute of Technology, 4800 Oak Grove Drive, Pasadena CA 91109, USA}

\author{S. Bertone}
\email{stefano.bertone@aiub.unibe.ch}
\affiliation{Observatoire de Paris, SYRTE, CNRS/UMR 8630, UPMC, 61 avenue de l'Observatoire, F-75014 Paris, France}
\affiliation{INAF, Astrophysical Observatory of Torino, University of Torino, Via Osservatorio 20, 10025 Pino Torinese (Torino), Italy }

\author{C. Le Poncin-Lafitte}
\email{christophe.leponcin@obspm.fr}
\affiliation{Observatoire de Paris, SYRTE, CNRS/UMR 8630, UPMC, 61 avenue de l'Observatoire, F-75014 Paris, France}

\date{\today}

\begin{abstract}
Given the extreme accuracy of modern space science, a precise relativistic modeling of observations is required. In particular, it is important to describe properly light propagation through the Solar System. For two decades, several modeling efforts based on the solution of the null geodesic equations have been proposed but they are mainly valid only for the first order Post-Newtonian approximation. However, with the increasing precision of ongoing space missions as Gaia, GAME, BepiColombo, JUNO or JUICE, we know that some corrections up to the second order have to be taken into account for future experiments. We present a procedure to compute the relativistic coordinate time delay, Doppler and astrometric observables avoiding the integration of the null geodesic equation. This is possible using the Time Transfer Function formalism, a powerful tool providing key quantities such as the time of flight of a light signal between two point-events and the tangent vector to its null-geodesic. Indeed we show how to compute the Time Transfer Functions and their derivatives (and thus range, Doppler and astrometric observables) up to the second post-Minkowskian order. We express these quantities as quadratures of some functions that depend only on the metric and its derivatives evaluated along a Minkowskian straight line. This method is particularly well adapted for numerical estimations. As an illustration, we provide explicit expressions in static and spherically symmetric space-time up to second post-Minkowskian order. Then we give the order of magnitude of these corrections for the range/Doppler on the BepiColombo mission and for astrometry   in a GAME-like observation.
\end{abstract}


\maketitle

\section{Introduction}

\par During the last twenty years, space science has made stunning progress. Indeed the accuracy on the tracking of probes increased drastically. For example, the Cassini spacecraft reached the level of few meters for the range and $3 \times 10^{-6} \,\rm{m/s}$ for the Doppler~\citep{iess:2007ve,bertotti:2003uq,kliore:2004zr}. In the near future, the BepiColombo mission should reach an accuracy of $10 \,\rm{cm}$ and $10^{-6}\, \rm{m/s}$ for range and Doppler respectively~\citep{milani:2002vn,iess:2009fk}. On the other hand, within the next years, Gaia's astrometric catalogue is expected to get positions, parallaxes and proper motions of a billion celestial objects with a precision of several microarcseconds \citep{2002EAS.....2.....B}, improving by a factor 1000 what was accomplished with HIPPARCOS \citep{2007ASSL..350.....V}. However we know that these high precision observations need to be reduced and interpreted in a complete relativistic framework \citep{2003AJ....126.2687S}. Several key points need to be considered, in particular a precise modeling for the propagation of the observed signal. In the limit of geometrical optics, it is well known that light rays follow null geodesics. Then radioscience (range \& Doppler) and astrometric observables are traditionally analyzed by determining the full light trajectory, by solving the null geodesic equations. This method works quite well within the first post-Newtonian (1PN) and post-Minkowskian (1PM) approximations, as it is shown by the results obtained in \cite{bertotti:1992rq,kopeikin:1997ys,kopeikin:1999kl,blanchet:2001ud,klioner:2003hs,kopeikin:2006ly, kopeikin:2007vn, crosta:2010qf}. In the context of the Schwarzschild-like geometry, a solution of the null geodesic equations has been derived at the post-post-Minkowskian (2PM) order in \citep{klioner:2009vn,klioner:2010fk}. The case of a static, spherically symmetric space-time has also been considered in \cite{ashby:2010fk} where a solution of the eikonal equation is found.

\par However, finding an analytical solution of the geodesic equations is a challenging task that requires complex calculations, in particular when one has to take into account the presence of mass multipoles and/or the effects due to the planetary motions. Moreover calculations become quite complicated in the 2PM approximation \cite{richter:1983oq} especially when space-time is not stationary \cite{brugmann:2005zr}. Nevertheless, it has been recently demonstrated that this task is not at all mandatory and can be replaced by another approach, initially based on the Synge World Function \citep{le-poncin-lafitte:2004cr} and then on the Time Transfer Functions (TTF)~\citep{teyssandier:2008nx}.

\par Indeed, within the TTF formalism, the solution of the null geodesic equation is advantageously replaced by the determination of the TTF and its first derivatives. In general, the determination of the TTF is as challenging as the integration of the null geodesic equations. Nevertheless, this task is really easier in a weak gravitational field. In particular, an algorithmic method to compute a PM expansion of the TTF at any order has been presented in \citep{teyssandier:2008nx}, the determination of the TTF being done by performing integrals of some functions of the space-time metric evaluated along a Minkowskian segment between the emitter and the receiver of the signal. Moreover from a computational point of view, the quadrature of a function taken along a straight line is easier than the full determination of the photon trajectory, which is a boundary value problem \cite{san-miguel:2007hc}.

\par In this article, we take advantage of the properties of the TTF formalism in order to construct a straightforward modeling of radioscience and astrometric observables. We present here a method to compute the TTF and its derivatives (and therefore coordinate time delay, frequency shift and astrometric observables) at 2PM order. Our method is particularly well adapted for numerical computation of radioscience and astrometric observables from the space-time metric. It can be used in General Relativity as well as in any alternative theories of gravity where the light propagation is described by the null geodesic equations at the geometric optic approximation, improving what two of us presented up to 1PM order in \cite{hees:2012fk}. 

\par The paper is organized as follows. In section~\ref{sec:not}, we present the notations and the conventions used through this paper. In section~\ref{sec:ttf}, we introduce briefly the TTF formalism and recall how to determine the TTF when the emitter and the receiver of the light ray are in motion. In section~\ref{sec:dop_astr}, we present a straightforward modeling of the radioscience and astrometric observables from the TTF. At this point, no expansion nor approximation is made and the observables are expressed in terms of the TTF and its first derivatives. In Section~\ref{sec:der_ttf}, we show how to compute the TTF and its derivatives up to 2PM order. In section~\ref{sec:applications}, we specify our formulas for a static, spherically symmetric space-time and apply them to a Schwarzschild-like geometry. We compute the order of magnitude of the 2PM terms in two cases. First we compute the values of the range and Doppler for BepiColombo and compare our results with those obtained in \citep{tommei:2010uq}. Second, considering a GAME-like observation \citep{gai:2012fk}, we simulate absolute and relative astrometric observations near the limb of the Sun to put in evidence the 2PM contribution to light deflection and aberration. In section~\ref{sec:conclusions}, we give our conclusions.

\section{Notation and conventions}  \label{sec:not}
In this paper $c$ is the speed of light in a vacuum and $G$ is the Newtonian gravitational constant. The Lorentzian metric of space-time $V_4$ is denoted by $g$. The signature adopted for $g$ is $(+---)$. We suppose that space-time is covered by some global quasi-Galilean coordinate system $(x^\mu)=(x^0,\bx )$, where $x^0=ct$, $t$ being a time coordinate, and $\bx=(x^i)$. We assume that the curves of equation $x^i$ = constants are timelike, which means that $g_{00}>0$ anywhere. We employ the vector notation $\ba$ in order to denote $(a^1,a^2,a^3)=(a^i)$. 
Considering two such quantities $\ba$ and $\bb$ we use $\ba \cdot \bb$ to denote $a^ib^i$ (Einstein convention on repeated indices is used). The quantity $\vert \ba \vert$ stands for the ordinary Euclidean norm of $\ba$. 
For any quantity $f(x^{\lambda})$, $f_{, \alpha}$ denotes the partial derivative of $f$ with respect to $x^{\alpha}$. 
In this paper, we are dealing with post-Minkowskian (PM) expansions. We suppose each quantity can be represented as a series in ascending power of $G$. The indices in parentheses characterize the order of perturbation. They are set up or down, depending on the convenience. For example, the space-time metric can be expanded as
\begin{equation}\label{eq:pm}
	g_{\mu\nu}=\eta_{\mu\nu}+\sum_{n=1}^\infty  g_{\mu\nu}^{(n)}
\end{equation}
	where $g_{\mu\nu}^{(n)}$ is of the order $\mathcal{O}(G^n)$. 
	
\section{Time Transfer Function formalism}\label{sec:ttf}

Let us consider two observers ${\cal O}_{\cal A}$ and ${\cal O}_{\cal B}$ located at point $\bx_A$ and $\bx_B$, respectively. We suppose that the past null cone at a given point $x_B=(ct_B,\bx_B)$ intersects the world line $\bx=\bx_A$ at only one point $x_A=(ct_A,\bx_A)$ (see Fig.~\ref{fig:gen}). The difference $t_B-t_A$ is the coordinate travel time of a light ray connecting the emission point $x_A$ with the reception point $x_B$. This quantity may be written as a Time Transfer Function~\cite{le-poncin-lafitte:2004cr,le-poncin-lafitte:2008fk,jaekel:2005zr,jaekel:2006uq}
\begin{equation}\label{eq:ttf}
	t_B-t_A=\mathcal T_r(\bx_A,t_B,\bx_B)=\mathcal T_e(t_A, \bx_A, \bx_B)\, ,
\end{equation}
where $\mathcal T_r$ and $\mathcal T_e$ are the time transfer functions (TTF) at reception and at emission, respectively. In the following, we consider only the case of the TTF at reception, but the discussion can be done in the same way by using the TTF at emission. TTF directly gives the coordinate propagation time of an electromagnetic signal and is therefore closely related to the Range observable~\cite{hees:2012fk}. The determination of the TTF is as challenging as the integration of the null geodesic equation~\cite{kopeikin:1999kl} but, in the weak field approximation, a general PM expansion of the TTF has been presented in~\citep{teyssandier:2008nx} which will be used in section~\ref{sec:der_ttf} to derive explicit equations up to the 2PM order. Generally speaking, neither the emitter ${\cal O}_{\cal A}$ nor the receiver ${\cal O}_{\cal B}$ of an electromagnetic signal are static. Instead, they are following a trajectory $\bx_{A}(t)$ and $\bx_{B}(t)$ usually parametrized by a coordinate time $t$. In this case, Eq.~(\ref{eq:ttf}) becomes an implicit relation since $\bx_A$ depends on $t_A$. In the weak field approximation, Eq.~(\ref{eq:ttf}) must then be read as 
\begin{equation}\label{eq:ttf_moving}
		t_B-t_A=\mathcal T_r(\bx_A(t_A),t_B,\bx_B(t_B))=\frac{\left|\bx_B(t_B)-\bx_A(t_A)\right|}{c}+\frac{1}{c}\Delta_{r}(\bx_A(t_A), t_B,\bx_B(t_B))\, ,
\end{equation}
with $\Delta_r / c$ the so called {\it delay function}~\citep{teyssandier:2008nx}. From an experimental point of view, the position of the emitter ${\cal O}_{\cal A}$ may be recorded at the time of emission $t_B$ rather than at the time of reception $t_A$, {\it i.e.} we may have more direct access to $x_A(t_B)$ rather than $x_A(t_A)$. Two approaches are possible. First an analytical solution can be derived by following the procedure presented in \cite{petit:2005uq}. For any quantity $Q_A(t)$ defined along the worldline of the observer ${\cal O}_{\cal A}$, let us put $\widetilde{Q}_A=Q_A(t_B)$. Thus we may write $\widetilde{\bx}_A$ for ${\bx}_A(t_B)$, $\widetilde{r}_A$ for $r_A(t_B)$, etc. The idea is now to expand the position of ${\cal O}_{\cal A}$ recorded at time $t_A$ with respect to coordinate time $t_B$ by a Taylor expansion as follow
\begin{equation}
	\bx_A(t_A)=	\tilde \bx_A+(t_A-t_B)\tilde\bv_A+\frac{1}{2}(t_A-t_B)^2\tilde \ba_A+\frac{1}{6}(t_A-t_B)^3
\tilde\bb_A+\dots 
\end{equation}
where $\tilde \bv_A =\bv_A(t_B)=\left.\frac{d\bx_A}{dt}\right|_{t_B}$, $\tilde \ba_A =\ba_A(t_B)=\left.\frac{d^2\bx_A}{dt^2}\right|_{t_B}$ and $\tilde \bb_A =\bb_A(t_B)=\left.\frac{d^3\bx_A}{dt^3}\right|_{t_B}$. The introduction of this expansion in (\ref{eq:ttf_moving}) leads to
\begin{eqnarray}
	&&t_B-t_A=\frac{\tilde D_{AB}}{c}+(t_B-t_A) \frac{\tilde \bv_A \cdot \tilde\bD_{AB}}{c\tilde D_{AB}} +\frac{(t_B-t_A)^2}{2c\tilde D_{AB}}\left[\tilde v_A^2-\tilde \ba_A\cdot \tilde\bD_{AB}-\left(\frac{\tilde\bv_A\cdot\tilde\bD_{AB}}{\tilde D_{AB}}\right)^2\right]  \nonumber\\
	&&+\frac{(t_B-t_A)^3}{2c\tilde D_{AB}}\left[\frac{1}{3}\tilde\bb_A\cdot\tilde\bD_{AB} - \tilde \bv_A\cdot\tilde\ba_A - \frac{\tilde v_A^2 \tilde \bv_A\cdot \tilde\bD_{AB}}{\tilde D_{AB}^2}+ \frac{(\tilde \bv_A\cdot \tilde\bD_{AB})(\tilde \ba_A\cdot \tilde\bD_{AB})}{\tilde D^2_{AB}} + \frac{(\tilde\bv_A\cdot\tilde\bD_{AB})^3}{\tilde D^4_{AB}}\right] +\frac{1}{c}\Delta_r(\tilde \bx_A,t_B,\bx_B)\nonumber\\
	&& -\frac{(t_B-t_A)}{c}  \frac{\partial\Delta_r(\tilde \bx_A,t_B,\bx_B)}{\partial x^i_A}\tilde v^i_A +\frac{(t_B-t_A)^2}{c}\left[\frac{1}{2}\frac{\partial\Delta_r(\tilde \bx_A,t_B,\bx_B)}{\partial x^i_A}\tilde a^i_A+\frac{\partial^2\Delta_r(\tilde \bx_A,t_B,\bx_B)}{\partial x^i_A\partial x^j_A}\tilde v^i_A\tilde v^j_A\right]+\dots \label{tototof1}
\end{eqnarray}
where $\tilde\bD_{AB}=\bx_B(t_B)-\bx_A(t_B)$ and $\tilde D_{AB}=\left|\tilde\bD_{AB}\right|$. An iterative solution of Eq. (\ref{tototof1}) gives
\begin{eqnarray}
\mathcal T_r(\bx_A(t_A),t_B,\bx_B(t_B))&=&\frac{\tilde D_{AB}}{c}+\frac{\tilde \bv_A\cdot \tilde\bD_{AB}}{c^2}+\frac{\tilde D_{AB}}{2c^3}\left[\tilde v_A^2+\left(\frac{\tilde\bv_A\cdot\tilde\bD_{AB}}{\tilde D_{AB}}\right)^2-\tilde\ba_A\cdot\tilde\bD_{AB}\right]+\nonumber\\
	&&\frac{1}{c^4}\left[(\tilde\bv_A \cdot \tilde\bD_{AB})(\tilde v_A^2-\tilde\ba_A\cdot\tilde\bD_{AB})+\frac{1}{6}\tilde D^2_{AB}\tilde\bb_A\cdot\tilde\bD_{AB}-\frac{1}{2}\tilde D^2_{AB}\tilde\bv_A\cdot\tilde\ba_A\right]+\label{eq:sagnac}\\
&&	\frac{1}{c}\Delta_r(\tilde \bx_A,t_B,\bx_B)- \frac{\tilde D_{AB}}{c^2}  \frac{\partial\Delta_r(\tilde \bx_A,t_B,\bx_B)}{\partial x^i_A}\tilde v^i_A  + \frac{\tilde\bv_A\cdot\tilde\bD_{AB}}{c^2 \tilde D_{AB}}\Delta_r(\tilde \bx_A,t_B,\bx_B)+\mathcal O(1/c^5)\nonumber.
\end{eqnarray}
Eq.~(\ref{eq:sagnac}) is a post-Newtonian (PN) formula since the TTF is expanded in terms of quantities such as $\tilde v_A/c$, $(\tilde \bD_{AB}\cdot \tilde\ba)/c^2$ that should be small in order to assure the convergence of this series. It should be noted that for this PN expansion $\Delta_r^{(1)}$ is considered of order $G/c^2$. This computation can be continued to higher orders if necessary. This analytical expansion includes  what is usually referred to as Sagnac-like terms \cite{blanchet:2001ud,2002PhRvD..66b4045L}. However, this expansion has the disadvantage to be valid for small velocities/accelerations only, which is not problematic in the Solar System but can be limiting in other applications like binary pulsars. Moreover, in this expansion, derivatives of $\Delta_r$ appear (and higher derivatives appear at higher orders) and these terms can become difficult to compute.
\vspace{0.5cm}
\par That is why a second approach, based on a numerical iterative process, is more practical. This procedure is standard and can be written as
\begin{subequations}\label{eq:iterative}
\begin{eqnarray}
	\textrm{Start: } && t_A^{(0)}=t_B-\mathcal T_r(\bx_A(t_B),t_B,\bx_B(t_B)) \\
	\textrm{Loop: } && t_A^{(i+1)}=t_B-\mathcal T_r(\bx_A(t_A^{(i)}),t_B,\bx_B(t_B)) \\
	\textrm{End: } && \textrm{when} \left|t_A^{(i+1)}-t_A^{(i)}\right|<\varepsilon
\end{eqnarray}
\end{subequations}
with $\varepsilon$ the desired accuracy. Each step of this iterative procedure requires one to evaluate the TTF. In practice, at least for Solar System applications, this procedure converges very quickly after two or three iterations. The main advantages of this procedure are that no PN approximation is done and it is easy to implement. 
\vspace{0.5cm}
\par These two procedures allow us to compute $t_A$, the coordinate emission time of the signal emitted along the world line $\bx_A(t)$, from the reception coordinate time $t_B$ and the coordinate of the receiver $\bx_B$. The analytical expansion (\ref{eq:sagnac}) is a PN expansion of $t_A$ up to ${\cal O} (1/c^4)$ while the iterative procedure (\ref{eq:iterative}) is valid up to any order.

\section{Doppler and astrometric observables from the Time Transfer Function}\label{sec:dop_astr}
\begin{figure}[hbt]
\begin{center}
\includegraphics[width=0.7\textwidth]{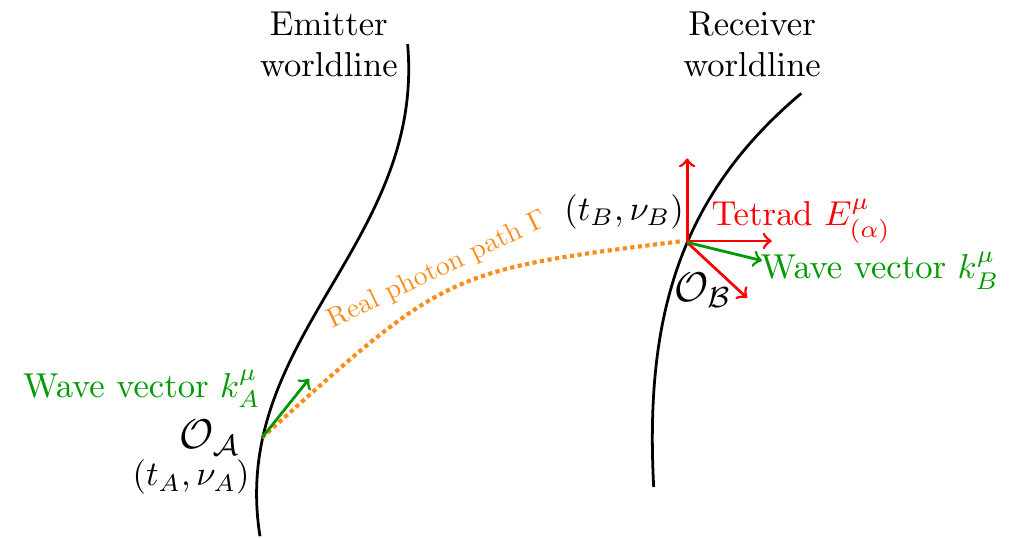}
\end{center}
\caption{Representation of the general geometry studied in this paper: a light signal of frequency $\nu_A$ is emitted by ${\cal O}_{\cal A}$ with a wave 4-vector of components $k_B^\mu$ and received by ${\cal O}_{\cal B}$ at a frequency $\nu_B$ and with a wave 4-vector of components $k_B^\mu$.}
\label{fig:gen}
\end{figure}

Here, we derive exact relativistic formulas to model the Doppler and astrometric observables as functions of the TTF and its partial derivatives. We consider three different observers ${\cal O}_{\cal A}$, ${\cal O}_{\cal A'}$ and ${\cal O}_{\cal B}$. We assume that ${\cal O}_{\cal A}$ and ${\cal O}_{\cal A'}$ are emitting light rays at coordinates $(t_A,\bx_A)$ and $(t_{A'},\bx_{A'})$, respectively. We assume also that these signals are received by ${\cal O}_{\cal B}$ at coordinates $(t_B,\bx_B)$.  ${\cal O}_{\cal B}$ is equipped with a comoving tetrad of components $E$. Fig.~\ref{fig:gen} illustrates the specific case of a light ray of frequency $\nu_A$ emitted by ${\cal O}_{\cal A}$ with a wave 4-vector of components $k_A^\mu$ and received by ${\cal O}_{\cal B}$ at a frequency $\nu_B$ and with a wave 4-vector of components $k_B^\mu$.
\subsection{Frequency shift observables} \label{sec:1wayfreqshift}
\par First, we focus on the one-way frequency shift between ${\cal O}_{\cal A}$ and ${\cal O}_{\cal B}$. Let us define it as follow
\begin{equation}
\label{FreqOneWay}
\left . \frac{\Delta \nu}{\nu}\right \vert_\textrm{A$\rightarrow$ B}^\textrm{one-way}= \frac{\nu_B}{\nu_A} - 1\, .
\end{equation}

\noindent It is well-known that the ratio $\nu_B/\nu_A$ can be expressed as \citep{synge:1960zr,teyssandier:2008fk}
\begin{equation}     \label{eq:nubnua}
\frac{\nu_B}{\nu_A}=\frac{u_B^\mu k^{B}_\mu}{u_A^\nu k^{A}_\nu} = \frac{u_B^0}{u_A^0}\frac{k^{B}_0}{k^{A}_0} \frac{1 + \beta_B^i \hat{k}^{B}_i}{1 + \beta_A^i \hat{k}^{A}_i} \, ,
\end{equation}
where  $u_{A}^\mu=(dx^\mu/ds)_{A}$ and $u_{B}^\mu=(dx^\mu/ds)_{B}$ are the four-velocity of ${\cal O}_{\cal A}$ and ${\cal O}_{\cal B}$, $\beta^i_{A}=dx_A^i/cdt$ and $\beta^i_{B}=dx_B^i/cdt$ are their coordinate velocities, $\hat{k}_i^A = \left(k^A_{i}/k^A_{0}\right)$ and $\hat{k}_i^B = \left(k^B_{i}/k^B_{0}\right)$, while $k^{A}_\mu$ and $k^{B}_\mu$ are the wave vectors tangents to the light ray at the point of emission $x_A$ and at the point of reception $x_B$, respectively.

\par The TTF formalism provides a direct way of defining the ratio of the spatial and temporal covariant components of the tangent vector to a photon trajectory $k^\mu = dx^\mu / d\sigma$, $\sigma$ being an affine parameter, at ${\cal O}_{\cal A}$ and ${\cal O}_{\cal B}$ as \cite{le-poncin-lafitte:2004cr} 
\begin{subequations}\label{eq:deflection_functions}
\begin{eqnarray}  
& &\left(\hat{k}_i\right)_A = \left(\frac{k_i}{k_0}\right)_A = \, c \, \frac{\partial {\cal T}_{r}}{\partial x^{i}_{A}} =  -N^i_{AB}+\frac{\partial \Delta_r}{\partial x^{i}_{A}}\, , \label{2d2} \\
& & \nonumber \\
& &\left(\hat{k}_i\right)_B = \left(\frac{k_i}{k_0}\right)_B =- c \, \frac{\partial  {\cal T}_{r}}{\partial x^{i}_{B}} \left[1 - \frac{\partial  {\cal T}_{r}} {\partial t_B}\right]^{-1}=-\left(N^i_{AB}+\frac{\partial \Delta_r}{\partial x^{i}_{B}}\right)\times\left[1 - \frac{1}{c}\frac{\partial  \Delta_{r}} {\partial t_B}\right]^{-1}\,  , \label{2d1} \\
& & \nonumber \\
& &\frac{(k_{0})_B}{(k_{0})_A} =  \, 1 - \frac{\partial  {\cal T}_{r}}{\partial t_{B}}  = 1 - \frac{1}{c}\frac{\partial  \Delta_{r}} {\partial t_B}\, , \label{2d3}
\end{eqnarray} 
\end{subequations}
\noindent where $N^i_{AB}=\dfrac{R_{AB}^i}{R_{AB}}$ with $R_{AB}^i=x_B^i-x_A^i$ and $R_{AB}=\left|\bx_B-\bx_A\right|$. Noting that 
\begin{equation}\label{eq:dtau}
u_{A/B}^0= \left[ g_{00}+2g_{0i} \beta^i+g_{ij} \beta^i\beta^j \right]^{-1/2}_{A/B} \, ,\end{equation} 
it is then straightforward to define the one-way frequency shift~(\ref{FreqOneWay}) as a function of $\Delta_{r}$ and its partial derivatives. Substituting for $\hat k_i$ from Eq.~\eqref{eq:deflection_functions} and inserting it in relation~\eqref{eq:nubnua} with using~\eqref{eq:dtau} , one gets the exact expression~\citep{teyssandier:2008fk,teyssandier:2009uq,hees:2012fk}
\begin{equation}\label{eq:doppler}
	\frac{\nu_B}{\nu_A}= \frac{\left[ g_{00}+2g_{0i} \beta^i +g_{ij} \beta^i\beta^j \right]^{1/2}_A}{\left[ g_{00}+2g_{0i} \beta^i +g_{ij} \beta^i\beta^j \right]^{1/2}_B}\times	 \frac{1-N^i_{AB} \beta^i_B -\beta^i_B \frac{\partial \Delta_r}{\partial x^i_B}-\frac{1}{c}\frac{\partial \Delta_r}{\partial t_B}}{1-N^i_{AB} \beta^i_A + \beta^i_A \frac{\partial \Delta_r}{\partial x^i_A}}.
\end{equation}
\par This modeling can be extended easily to a multi-way frequency shift. For example, let us consider a signal emitted with a frequency $\nu_A$ by an observer ${\cal O}_{\cal A}$, transmitted by an observer ${\cal O}_{\cal B}$ and then received by an observer ${\cal O}_{\cal C}$ at a frequency $\nu_C$, which can eventually be ${\cal O}_{\cal A}$ for a 2-way frequency shift.  The frequency shift between ${\cal O}_{\cal A}$ and ${\cal O}_{\cal C}$ is defined in the same way as for the 1-way 
\begin{equation}\label{eq:freq_multi}
	\left . \frac{\Delta \nu}{\nu}\right \vert_\textrm{A$\rightarrow$ C}= \frac{\nu_C}{\nu_A} - 1\, .
\end{equation}
The ratio $\nu_C/\nu_A$ can be decomposed as follows
\begin{equation} \label{eq:partialfreqshift}
	 \frac{\nu_C}{\nu_A}=\frac{\nu_C}{\nu_{B,e}}\delta\nu_B\frac{\nu_{B,r}}{\nu_A}\, ,
\end{equation}
where $\nu_{B,r}$ is the proper frequency received by the observer ${\cal O}_{\cal B}$, while $\nu_{B,e}$ is the proper frequency emitted by the same observer. The factor $\delta\nu_B\equiv \nu_{B,e}/\nu_{B,r}$ stands for any frequency shift, $i.e.$ due to a transponder, introduced between the reception and re-emission of the signal. The computation of the multi-ways frequency shift is straightforward: the two terms $\nu_C/\nu_{B,e}$ and $\nu_{B,r}/\nu_A$ from (\ref{eq:partialfreqshift}) are 1-way frequency shifts and can be computed using (\ref{eq:doppler}). 
This procedure can be generalized easily if more links are needed.

\subsection{Astrometric observables} \label{sec:astroobs}

\par The goal of astrometry is to determine the position of celestial bodies from angular observations. We focus on two main approaches. First, we consider the modeling of the direction of incidence of a light ray in a given reference frame, which gives an absolute positioning of the studied object on a celestial sphere. Second, we consider the case of the angular separation of two light sources.

\par One way to get a covariant definition of the absolute positioning of a light source is to use the tetrad formalism~\citep{brumberg:1991uq,misner:1973fk,weinberg:1972vn,klioner:1992ly} thus giving the direction of observation of an incoming light ray in a tetrad $E$ comoving with the observer $\mathcal {O_B}$ (see Fig.~\ref{fig:gen}). Let us note $E^\mu_{\left<\alpha\right>}$ the components of this tetrad, where $\left<\alpha\right>$ corresponds to the tetrad index and $\mu$ is a normal tensor index that can be lowered and raised by use of the metric. The tetrad is assumed to be orthonormal so that
\begin{equation}
g_{\mu\nu}	E^\mu_{\left<\alpha\right>}	E^\nu_{\left<\beta\right>}=\eta_{\left<\alpha\right>\left<\beta\right>}.
\end{equation}
Vector $E^\nu_{\left<0\right>}$ is chosen unit and timelike, and consequently $E^\nu_{\left<i\right>}$ are unit and spacelike. The components of the tetrad allow us to transform the coordinates of the wave vector from the global coordinate frame to the tetrad frame
\begin{equation}\label{eq:transf_k}
	k_{\left<\alpha\right>}=E^\mu_{\left<\alpha\right>}k_\mu
\end{equation}
where $k_{\mu}$ are the coordinates of the wave vector in the global frame (represented on Fig.~\ref{fig:gen}) while $k_{\left<\alpha\right>}$ are the coordinates of the same vector in the tetrad frame. The incident direction of the light ray in the tetrad frame (which is a relativistic observable) is given by the normalization
\begin{equation} \label{eq:normincdir}
	n^{\left< i\right>}=\frac{k^{\left<i \right>}}{\sqrt{\delta_{jk}k^{\left<j\right>}k^{\left<k \right>}}} =\frac{k^{\left< i\right>}}{k^{\left< 0 \right>}}=-\frac{k_{\left< i \right>}}{k_{\left< 0 \right>}} \; ,
\end{equation}
where we used the properties of the null-vector $k^{\left< i \right>}$ and the fact that the metric tensor has a Minkowskian form in the tetrad frame. Using the transformation law~(\ref{eq:transf_k}) into Eq.~(\ref{eq:normincdir}), one gets
\begin{equation}
	n^{\left< i \right>}=-\frac{E^0_{\left< i \right>}k_0+E^j_{\left< i \right>}k_j}{E^0_{\left< 0 \right>}k_0+E^j_{\left< 0 \right>}k_j }=-\frac{E^0_{\left< i \right>}+E^j_{\left< i \right>}\hat k_j}{E^0_{\left< 0 \right>}+E^j_{\left< 0 \right>}\hat k_j } \; ,
\end{equation}
where $\hat k_j$ are the deflection functions at ${\cal O}_{\cal B}$ defined in (\ref{2d1}). This expression is consistent with the one derived in~\cite{klioner:2004fk}. Using the relation (\ref{2d1}) one can then express the incoming direction of the light ray in terms of the reception delay function and its derivatives~\citep{bertone:2012fk,bertone:2014a} as
\begin{equation}\label{eq:astro}
	n^{\left< i \right>}
=  -\frac{E^0_{\left< i \right>}\left(1-\frac{1}{c}\frac{\partial \Delta_r}{\partial t_B}\right)-E^j_{\left< i \right>}N^j-E^j_{\left< i \right>}\frac{\partial\Delta_r}{\partial x^j_B}}{E^0_{\left< 0 \right>}\left(1-\frac{1}{c}\frac{\partial \Delta_r}{\partial t_B}\right) -E^j_{\left< 0 \right>}N^j-E^j_{\left< 0 \right>}\frac{\partial\Delta_r}{\partial x^j_B}} \; ,
\end{equation}
which is an exact formula.

\vspace{0.5cm}

 Let us now examine the second kind of astrometric observations, namely the modeling of angular distance between two celestial bodies. This observable can also be computed within the TTF formalism. We assume that two different light sources ${\cal O}_{\cal A}$ and ${\cal O}_{\cal A'}$ are emitting a light ray $\Gamma$ and $\Gamma'$, respectively. These light rays are received simultaneously by ${\cal O}_{\cal B}$ at coordinates $(t_B,{\bm x}_B)$. We denote by $k$ and $k'$ the wave vector of $\Gamma$ and $\Gamma'$ at ${\cal O}_{\cal B}$, respectively. Using expression (\ref{2d1}), we construct the ratio $\left(\hat{k}_j\right)_B$ corresponding to $\Gamma$ and $\left(\hat{k}_j'\right)_B$ describing $\Gamma'$, which require an expression for the derivatives of the TTF whose expression up to the 2PM order will be given in section~\ref{sec:der_ttf}. It is straightforward to show that the angular distance $\phi$ between ${\cal O}_{\cal A}$ and ${\cal O}_{\cal A'}$, as observed by a moving observer $\cal O_B$, can be written as \citep{teyssandier:2006fk} 
\begin{equation}\label{eq:ang_sep}
	\sin^2 \frac{\phi}{2}=-\frac{1}{4}\left[\frac{\left(g_{00}+2g_{0k}\beta^k+g_{kl}\beta^k\beta^l\right)g^{ij}(\hat{k}'_i-\hat{k}_i)(\hat{k}'_j-\hat{k}_j)}{(1+\beta^m\hat k_m)(1+\beta^l\hat k_l')}\right]_B \, ,
\end{equation}
where $\beta^i_B=(dx^i/cdt)_B$ is the coordinate velocity of ${\cal O}_{\cal B}$ at coordinates $(t_B,{\bm x}_B)$.
\section{Post-Minkowskian expansion of the time transfer function and its derivatives} \label{sec:der_ttf}
In Section~\ref{sec:dop_astr}, we have presented a method to compute Doppler and astrometric observables in an exact form depending explicitly on the expression of the TTF and its derivatives. In this section, we present a way to derive these quantities up to 2PM order as integrals of some functions of the space-time metric taken along a straight line. In the weak field approximation, the expression of $\mathcal{T}_{r}$ as a formal PM series has been derived by \cite{teyssandier:2008nx} and can be written in ascending powers of $G$ as
\begin{equation}\label{eq:TTF_PM}
{\cal T}_{	r}(\bx_A, t_B, \bx_B)=\frac{R_{AB}}{c}+ \frac{1}{c}\sum_{n=1}^\infty\Delta^{(n)}_{r}(\bx_A, t_B,\bx_B)\, ,
\end{equation}
where $\Delta^{(n)}_{r}$ is of the order $\mathcal{O}(G^n)$. The goal of this section is then to derive analytical formulas for the delay functions $\Delta_r^{(1)}$, $\Delta_r^{(2)}$ and their derivatives \citep{hees:2012uq} up to 2PM order.

\subsection{Notations and variables used}\label{sec:notations}
In the following, we provide some useful notations used throughout this paper.
First of all, the Minkowskian path between the emitter and the receiver (which is a straight line) is parametrized by $\lambda$ (whose values are between 0 and 1) and is given by
\begin{subequations}\label{eq:z}
\begin{eqnarray}
	z^0(\lambda)&=&ct_B-\lambda R_{AB}\\
	\bz(\lambda)&=&\bx_B-\lambda \bR_{AB}=\bx_B(1-\lambda)+\lambda \bx_A.
\end{eqnarray}
\end{subequations}
We introduce the derivatives of these expressions with respect to the variables $\bx_{A/B}$, {\it i.~e.} the quantities
\begin{subequations}\label{eq:dz}
	\begin{eqnarray}
		z^0_{,(Ai)}(\lambda)&=&\frac{\partial z^0(\lambda)}{\partial x^i_A}=\lambda N^i_{AB}\label{eq:dz0dzi} \, ,\\
		z^0_{,(Bi)}(\lambda)&=&\frac{\partial z^0(\lambda)}{\partial x^i_B}=-\lambda N^i_{AB} \, , \\
		z^j_{,(Ai)}(\lambda)&=&\frac{\partial z^j(\lambda)}{\partial x^i_A}=\lambda\delta^j_i\label{eq:dzdxia} \, , \\
		z^j_{,(Bi)}(\lambda)&=&\frac{\partial z^j(\lambda)}{\partial x^i_B}=(1-\lambda)\delta^j_i \, , \\
		z^0_{,(Ak)(Al)}(\lambda)&=&\frac{\partial^2 z^0}{\partial x^l_A \partial x^k_{A}}=\frac{\lambda}{R_{AB}}(N^k_{AB}N^l_{AB}-\delta_{kl}) \, , \\
		z^0_{,(Ak)(Bl)}(\lambda)&=&\frac{\partial^2 z^0}{\partial x^l_B \partial x^k_{A}}=-\frac{\lambda}{R_{AB}}(N^k_{AB}N^l_{AB}-\delta_{kl}) \, .
	\end{eqnarray}
\end{subequations}
We will use the functions $p$ and $p_{(n)}$ defined from the PM expansion of the space-time metric as follows
\begin{subequations}\label{eq:p}
	\begin{eqnarray}
p_{(n)}(\lambda)&=&p\left[g^{\mu\nu}_{(n)}(z^\beta(\lambda)),N^i_{AB},R_{AB}\right]=\frac{R_{AB}}{2}\left[g^{00}_{(n)}-2N^k_{AB}g^{0k}_{(n)}+N^k_{AB}N^l_{AB}g^{kl}_{(n)}\right]_{z^\beta(\lambda)}.
	\end{eqnarray}
We also define a similar expression with the metric replaced by its derivatives
	\begin{eqnarray}
p_{(n)\alpha}(\lambda)&=&p\left[g^{\mu\nu}_{(n),\alpha}(z^\beta(\lambda)),N^i_{AB},R_{AB}\right]=\frac{R_{AB}}{2}\left[g^{00}_{(n),\alpha}-2N^k_{AB}g^{0k}_{(n),\alpha}+N^k_{AB}N^l_{AB}g^{kl}_{(n),\alpha}\right]_{z^\beta(\lambda)} \; .\label{eq:p2}
	\end{eqnarray}
\end{subequations}
It is worth noticing that the last definition corresponds to the derivative of $p_{(n)}(\lambda)$ with respect to $z^\beta$, by keeping $N^i_{AB}$ and $R_{AB}$ constants
\begin{equation}\label{eq:pn_der}
	p_{(n)\alpha}(\lambda)=\left.\frac{\partial p_{(n)}(\lambda)}{\partial z^\alpha}\right|_{N^i_{AB}, R_{AB} \textrm{ cst}}.
\end{equation}

We will also use the functions $q^j_{(n)}$ that are defined by the derivative of $p_{(n)}$ with respect to $x^j_A$ by keeping $z^{\beta}$ constant 
\begin{subequations}\label{eq:q}
	\begin{eqnarray}
 		q^j_{(n)}(\lambda)&=& \left.\frac{\partial p_{(n)}}{\partial x^j_A}\right|_{z^\beta={\rm cst}}= \frac{1}{2}\left[-N^j_{AB}g^{00}_{(n)}+2g^{0j}_{(n)}-2g^{jk}_{(n)}N_{AB}^k+N_{AB}^kN_{AB}^lN^j_{AB} g^{kl}_{(n)}\right]_{z^\beta(\lambda)} \; .\label{eq:q1}
	\end{eqnarray}
It is then straightforward to show that
\begin{eqnarray}	
 q^j_{(n)}(\lambda)&=&-\left.\frac{\partial p_{(n)}}{\partial x^j_B}  \right|_{z^\beta={\rm cst}} \; .
\end{eqnarray}
We define a similar expression by replacing the metric by its derivatives
		\begin{eqnarray} 
		 q^j_{(n)\alpha}(\lambda)&=&			\left.\frac{\partial p_{(n)\alpha}}{\partial x^j_A}\right|_{z^\beta={\rm cst}}=-\left.\frac{\partial p_{(n)\alpha}}{\partial x^j_B}  \right|_{z^\beta={\rm cst}}  \nonumber\\
		&=&	\frac{1}{2}\left[-N^j_{AB}g^{00}_{(n),\alpha}+2g^{0j}_{(n),\alpha}-2g^{jk}_{(n),\alpha}N_{AB}^k+N_{AB}^kN_{AB}^lN^j_{AB} g^{kl}_{(n),\alpha}\right]_{z^\beta(\lambda)} \; .
	\end{eqnarray}
Finally, the derivatives of $q^j_{(n)}$ with respect to $x^k_A$ by keeping $z^\beta$ constant are denoted by	
	\begin{eqnarray}
           s^{ij}_{(n)}(\lambda)&=&\left.\frac{\partial q^i_{(n)}}{\partial x^j_A}  \right|_{z^\alpha={\rm cst}}=-\left.\frac{  q^i_{(n)}}{\partial x^k_B}  \right|_{z^\alpha={\rm cst}} \nonumber\\
		&=&\frac{1}{2R_{AB}}\left[g^{00}_{(n)}(\delta^{ij}-N_{AB}^iN_{AB}^j) + 2g^{ij}_{(n)}- 2 N_{AB}^k(g^{ik}_{(n)}N^j_{AB}+g^{jk}_{(n)}N^l_{AB}) \right. \\ 
		&& \qquad \qquad\qquad \qquad \left.+g^{kl}_{(n)}N^k_{AB}N^l_{AB}(3N^i_{AB}N^j_{AB}-\delta^{ij})\right]_{z^\alpha(\lambda)} \; .  \nonumber
	\end{eqnarray}
\end{subequations}

\subsection{Expansion at first PM order}
The expression of $\Delta_r^{(1)}$ is given in \citep{teyssandier:2008nx} as an integral taken along $z^\alpha(\lambda)$ of the components of the metric tensor 
\begin{equation}\label{eq:delta1_exp}
	\Delta_r^{(1)} (\bx_A,t_B,\bx_B)= \frac{R_{AB}}{2} \int_0^1  \left[g^{00}_{(1)} - 2N^i_{AB} g^{0i}_{(1)} + N^i_{AB} N_{AB}^j g^{ij}_{(1)}  \right]_{z^\alpha(\lambda)} d\lambda \, .
\end{equation}
Using the notations~(\ref{eq:p}), we rewrite Eq.~\eqref{eq:delta1_exp} as
\begin{equation}\label{eq:delta1}
	\Delta_r^{(1)} (\bx_A,t_B,\bx_B)=\int_0^1 p\left[g^{\mu\nu}_{(1)}(z^\beta\left(\lambda)\right),N^i_{AB},R_{AB}\right]d\lambda=\int_0^1 p_{(1)}(\lambda)d\lambda.
\end{equation}

The derivatives of $\Delta_r^{(1)}$ can then be computed from~(\ref{eq:delta1}) by inverting the integral and the partial derivative and by using the chain rules
\begin{equation} \label{eq:Delta1b}
	\frac{\partial{\Delta}^{(1)}_{r}}{\partial x^{i}_{A}} (\bx_A, t_B, \bx_B)=\int_0^1\left[\left.\frac{\partial p_{(1)}(\lambda)}{\partial z^\alpha}\right|_{N^i_{AB}, R_{AB} \textrm{ cst}}\frac{\partial z^\alpha}{\partial x^i_A} + \left.\frac{\partial p_{(1)}(\lambda)}{\partial x^i_A}\right|_{z^\beta={\rm cst}}\right]d\lambda.
\end{equation}
This can be rewritten using relations~(\ref{eq:dz}), (\ref{eq:pn_der}) and~(\ref{eq:q1}) as
\begin{subequations} \label{eq:dDelta1} 
\begin{equation}
		\frac{\partial{\Delta}^{(1)}_{r}}{\partial x^{i}_{A}} (\bx_A, t_B, \bx_B)=\int_0^1\left[ p_{(1)\alpha}(\lambda)z^\alpha_{,(Ai)}(\lambda)+q^i_{(1)}(\lambda)\right]d\lambda\; ,
\end{equation}
while a similar reasoning leads to
\begin{eqnarray}
		\frac{\partial{\Delta}^{(1)}_{r}}{\partial x^{i}_{B}} (\bx_A, t_B, \bx_B) &=& \int_0^1  \left[ p_{(1)\alpha}(\lambda) z^\alpha_{,(Bi)}(\lambda) - q^i_{(1)} (\lambda) \right] d\lambda\, , \\
	\frac{\partial{\Delta}^{(1)}_{r}}{\partial t_{B}} (\bx_A, t_B, \bx_B) &=&c \int_0^1   p_{(1)0}(\lambda)  d\lambda\, . 
\end{eqnarray} 
\end{subequations}
Eqs.~\eqref{eq:Delta1b}-\eqref{eq:dDelta1} are equivalent to those derived in \cite{hees:2012fk}. When replaced into Eq.~\eqref{eq:doppler}, Eq.~\eqref{eq:astro} and Eq.~\eqref{eq:ang_sep}, they give a full description of Doppler and astrometric observables at 1PM. 
\bigskip 

Some other quantities will be useful for the computations at the 2PM order. In particular, $\Delta_r^{(1)}(\bz(\lambda),t_B,\bx_B)$ is defined similarly to~(\ref{eq:delta1_exp}) as
\begin{equation}\label{eq:delta_z}
		\Delta_r^{(1)}(\bz(\lambda),t_B,\bx_B)=\frac{R_{zB}}{2} \int_0^1  \left[g^{00}_{(1)} - 2N^i_{zB} g^{0i}_{(1)} + N^i_{zB} N_{zB}^j g^{ij}_{(1)}  \right]_{y^\alpha(\mu)} d\mu=\int_0^1 p\left[g^{\mu\nu}_{(1)}(y^\alpha\left(\mu)\right),N^i_{zB},R_{zB}\right]d\mu
\end{equation}
where the integral is performed over a straight line joining $z^\alpha (\lambda)$ to $x^\alpha_B$. This path is parametrized by $y^\alpha(\mu)$, whose components are given by
\begin{subequations}
\begin{eqnarray}
	y^0(\mu)&=&ct_B-\mu R_{zB}=ct_B-\mu \left| \bx_B - \bz (\lambda)  \right| \\
	\by(\mu)&=&\bx_B-\mu \left(\bx_B-\bz(\lambda)\right) \; .
\end{eqnarray}
\end{subequations}
Using the definition of $z^\alpha$ given in Eq.~(\ref{eq:z}), the last expression becomes
\begin{equation}\label{eq:y_alpha}
	y^\alpha(\mu)=z^\alpha(\mu\lambda) \; .
\end{equation}
Inserting this relation in (\ref{eq:delta_z}) and noticing that 
\begin{subequations}\label{eq:eq_z_a}
	\begin{eqnarray}
		R_{zB}&=&\lambda R_{AB} \; 
	\end{eqnarray}
	and
	\begin{eqnarray}
		\bN_{zB}&=&\bN_{AB}\; , 
	\end{eqnarray}
\end{subequations}
one gets
$$
\Delta_r^{(1)}(\bz(\lambda),t_B,\bx_B)=\int_0^1 p\left[g^{\mu\nu}_{(1)}(z^\beta\left(\lambda\mu)\right),N^i_{AB},\lambda R_{AB}\right]d\mu \; .
$$
Since $p$ is linear with respect to $R_{AB}$, the last expression becomes
\begin{equation}\label{eq:Deltaz}
			\Delta_r^{(1)}(\bz(\lambda),t_B,\bx_B)=\lambda\int_0^1 p\left[g^{\mu\nu}_{(1)}(z^\beta\left(\lambda\mu)\right),N^i_{AB},R_{AB}\right]d\mu=\lambda\int_0^1 p_{(1)}(\lambda\mu)d\mu\;.
\end{equation}
We shall also need the quantity $\frac{\partial \Delta_r^{(1)}(\bz(\lambda),t_B,\bx_B)}{\partial x^i}$ (where the derivative is taken with respect to $\bz(\lambda)$). To compute it, we apply the chain rules to Eq.~(\ref{eq:delta_z}) so that we get
\begin{equation}\label{eq:dDz_dx_tmp}
	\frac{\partial \Delta_r^{(1)}(\bz(\lambda),t_B,\bx_B)}{\partial x^i}=\int_0^1\left[\left.\frac{\partial p\left[g^{\mu\nu}_{(1)}(y^\beta(\mu)),N_{zB}^i,R_{zb}\right]}{\partial y^\alpha}\right|_{N^i_{zB}, R_{zB} \textrm{ cst}}\frac{\partial y^\alpha}{\partial z^i} + \left.\frac{\partial p\left[g^{\mu\nu}_{(1)}(y^\beta(\mu)),N_{zB}^i,R_{zb}\right]}{\partial z^i}\right|_{y^\beta={\rm cst}}\right]d\mu.
\end{equation}
The first part of the integrand gives
\begin{equation}
\label{eq:dpdy}
\left.\frac{\partial p\left[g^{\mu\nu}_{(1)}(y^\beta(\mu)),N_{zB}^i,R_{zb}\right]}{\partial y^\alpha}\right|_{N^i_{zB}, R_{zB} \textrm{ cst}} =p\left[g^{\mu\nu}_{(1),\alpha}(y^\beta(\mu)),N_{zB}^i,R_{zb}\right]=\lambda p\left[g^{\mu\nu}_{(1),\alpha}(z^\beta(\lambda\mu)),N_{AB}^i,R_{Ab}\right]=\lambda p_{(1)\alpha}(\lambda\mu)\; ,
\end{equation}
while the derivative of $y^\alpha$ is given by
$$
\frac{\partial y^0(\mu)}{\partial z^i}=\mu N^i_{zB}=\mu N^i_{AB},  \qquad \qquad \frac{\partial y^j(\mu)}{\partial z^i}=\mu\delta^j_i.
$$
The comparison of these expressions with~(\ref{eq:dz0dzi}) and~(\ref{eq:dzdxia}) gives then
\begin{equation}\label{eq:dy}
\lambda \frac{\partial y^\alpha(\mu)}{\partial z^i}=\frac{\partial z^\alpha(\lambda\mu)}{\partial x^i_A}=z^\alpha_{,(Ai)}(\lambda\mu).
\end{equation}
The computation of the last quantity in the integrand is also straightforward. We get
\begin{equation} \label{eq:dpdzb}
\left.\frac{\partial p\left[g^{\mu\nu}_{(1)}(y^\beta(\mu)),N_{zB}^i,R_{zb}\right]}{\partial z^i}\right|_{y^\beta={\rm cst}}=\frac{1}{2}\left[-N^i_{zB}g^{00}_{(1)}+2g^{0i}_{(1)}-2g^{ik}_{(1)}N_{zB}^k+N_{zB}^kN_{zB}^lN_{zB}^i g^{kl}_{(1)}\right]_{y^\beta(\mu)}\; .
\end{equation}
Using (\ref{eq:eq_z_a}) and (\ref{eq:y_alpha}) into Eq.~\eqref{eq:dpdzb} and comparing to Eq.~\eqref{eq:q}, one can finally set
\begin{equation}\label{eq:dp_dz}
\left.\frac{\partial p\left[g^{\mu\nu}_{(1)}(y^\beta(\mu)),N_{zB}^i,R_{zb}\right]}{\partial z^i}\right|_{y^\beta={\rm cst}}=q_{(1)}^i(\lambda\mu).
\end{equation}
We can now use (\ref{eq:dpdy}), (\ref{eq:dy}) and (\ref{eq:dp_dz}) into (\ref{eq:dDz_dx_tmp}) to get
\begin{subequations}\label{eq:Dz}
\begin{equation}\label{eq:Dz_dx}
		\frac{\partial \Delta_r^{(1)}(\bz(\lambda),t_B,\bx_B)}{\partial x^i}=\int_0^1\left[p_{(1)\alpha}(\lambda\mu)z^\alpha_{,(Ai)}(\lambda\mu)+q_{(1)}^i(\lambda\mu)\right]d\mu \; .
\end{equation}

The next quantities of interest are the derivatives of $\Delta_r^{(1)}(\bz(\lambda),t_B,\bx_B)$ with respect to $x^i_{A/B}$ and $t_B$. Once again, we apply the chain rules to~(\ref{eq:Deltaz}) so that
\begin{eqnarray}
	\frac{\partial \Delta_r^{(1)}(\bz(\lambda),t_B,\bx_B)}{\partial x_A^i}&=&\lambda \int_0^1\left[ p_{(1)\alpha}(\lambda\mu)z^\alpha_{,(Ai)}(\lambda\mu)+q^i_{(1)}(\lambda\mu)\right]d\mu \; , \label{eq:dDz_dxia}\\
		\frac{\partial \Delta_r^{(1)}(\bz(\lambda),t_B,\bx_B)}{\partial x_B^i}&=&\lambda \int_0^1\left[ p_{(1)\alpha}(\lambda\mu)z^\alpha_{,(Bi)}(\lambda\mu)-q^i_{(1)}(\lambda\mu)\right]d\mu \; ,\label{eq:dDz_dxib}\\
			\frac{\partial \Delta_r^{(1)}(\bz(\lambda),t_B,\bx_B)}{\partial t_B}&=&c\; \lambda \int_0^1 p_{(1)0}(\lambda\mu)d\mu \label{eq:dDz_dtb} \; .
\end{eqnarray}
\end{subequations}

Finally, we explicit the second derivatives of $\Delta_r^{(1)}(\bz(\lambda),t_B,\bx_B)$ using the chain rules from Eq.~(\ref{eq:Dz_dx}) as
\begin{subequations}\label{eq:d2Delta}
\begin{eqnarray}
	\frac{\partial^2{\Delta}^{(1)}_{r}}{\partial x^{i}_{A}\partial x^{j}} (\bm z(\lambda),t_B,\bx_B)&=&\int_0^1 \left[ \right. p_{(1)\alpha \beta} z^\alpha_{,(Aj)} z^\beta_{,(Ai)} + q^i_{(1)\alpha} z^\alpha_{,(Aj)} + p_{(1)\alpha} z^\alpha_{,(Aj)(Ai)} +q^j_{(1)\alpha} z^\alpha_{,(Ai)} + s^{ji}_{(1)} \left. \right]_{\lambda \mu} d\mu \, , \\
		\frac{\partial^2{\Delta}^{(1)}_{r}}{\partial x^{i}_{B}\partial x^{j}} (\bm z(\lambda),t_B,\bx_B)&=&\int_0^1 \left[ \right. p_{(1)\alpha \beta} z^\alpha_{,(Aj)} z^\beta_{,(Bi)} - q^i_{(1)\alpha} z^\alpha_{,(Aj)} + p_{(1)\alpha} z^\alpha_{,(Aj)(Bi)} + q^j_{(1)\alpha} z^\alpha_{,(Bi)} - s^{ji}_{(1)} \left. \right]_{\lambda \mu} d\mu \, , \\
	\frac{\partial^2{\Delta}^{(1)}_{r}}{\partial t_{B}\partial x^{j}} (\bm z(\lambda),t_B,\bx_B)&=&c\int_0^1\left[p_{(1)\alpha 0}(\lambda\mu)z^\alpha_{,(Aj)}(\lambda\mu)+q^j_{(1)0}(\lambda\mu)\right]d\mu.
\end{eqnarray}
\end{subequations}

To summarize, the expansion of the TTF and its derivatives at 1PM order are given by (\ref{eq:delta1}) and (\ref{eq:dDelta1}). The quantities (\ref{eq:Deltaz}), (\ref{eq:Dz}) and \eqref{eq:d2Delta} will be useful in the calculation of the 2PM expansion of the TTF presented in the following.

\subsection{Expansion at second PM order}
The expression of $\Delta_r^{(2)}$ can also be derived from \citep{teyssandier:2008nx} and rewritten with our notations as
\begin{equation}\label{eq:delta2}
	\Delta_r^{(2)}(\bx_A,t_B,\bx_B) = \int_0^1 \left[ \mathcal I_1(\lambda) + \mathcal I_2 (\lambda) +\mathcal I_3(\lambda) \right] d\lambda\,
\end{equation}
with
\begin{subequations}
\begin{eqnarray}
	\mathcal I_1(\lambda) &=& p_{(2)} (\lambda) - \Delta_r^{(1)} (\bm z(\lambda),t_B,\bm x_B)\  p_{(1)0}(\lambda)  \nonumber\\ 
	&=&p_{(2)} (\lambda) - \lambda\; p_{(1)0}(\lambda)\int_0^1 p_{(1)}(\lambda\mu)d\mu \, ,
\end{eqnarray}
where we have used (\ref{eq:Deltaz}),
\begin{eqnarray}
	\mathcal I_2(\lambda) &=& \left[ R_{AB}\  \gab 0i1 - R^k_{AB}\  \gab ik1  \right]_{z^\alpha(\lambda)}  \times \frac{\partial \Delta_r^{(1)}}{\partial x^i}(\bz(\lambda),t_B,\bx_B)   \nonumber\\
	&=& \left[ R_{AB}\  \gab 0i1 - R^k_{AB}\  \gab ik1  \right]_{z^\alpha(\lambda)}\times \int_0^1 \left[ p_{(1)\alpha}(\lambda\mu) z^\alpha_{,(Ai)}(\lambda\mu) + q^i_{(1)}(\lambda\mu) \right] d\mu  \, ,
\end{eqnarray}
where we have used (\ref{eq:Dz_dx}) and
\begin{eqnarray}
	\mathcal I_3(\lambda) &=& - \frac{R_{AB}}{2} \sum_{j=1}^3 \left[ \frac{\partial \Delta_r^{(1)}}{\partial x^j}(\bz(\lambda),t_B,\bx_B) \right]^2 \nonumber\\
	&=&- \frac{R_{AB}}{2} \sum_{j=1}^3 \left\{\int_0^1 \left[ p_{(1)\alpha}(\lambda\mu) z^\alpha_{,(Aj)}(\lambda\mu) +  q^j_{(1)}(\lambda\mu) \right] d\mu\right\}^2 \, ,
\end{eqnarray}      
\end{subequations}
where we have used the relation (\ref{eq:Dz_dx}).

Applying extensively the chain rules, we can now derive the expression of the partial derivatives of Eq.~\eqref{eq:delta2} as
\begin{subequations}   \label{dDelta2} 
\begin{eqnarray}
	\frac{\partial{\Delta}^{(2)}_{r}}{\partial x^{i}_{A/B}} (\bx_A, t_B, \bx_B) &=& \int_0^1  \left[\frac{\partial \mathcal I_1}{\partial x^i_{A/B}}(\lambda) +\frac{\partial \mathcal I_2}{\partial x^i_{A/B}}(\lambda)+\frac{\partial \mathcal I_3}{\partial x^i_{A/B}}(\lambda)\right] d\lambda \\
\frac{\partial{\Delta}^{(2)}_{r}}{\partial t_B} (\bx_A, t_B, \bx_B) &=& \int_0^1  \left[\frac{\partial \mathcal I_1}{\partial t_B} (\lambda)+\frac{\partial \mathcal I_2}{\partial t_B}(\lambda)+\frac{\partial \mathcal I_3}{\partial t_B}(\lambda)\right] d\lambda \, ,
\end{eqnarray}    
\end{subequations}
where the derivatives can be written as follows 
\begin{subequations}\label{k2PM}
\begin{eqnarray} 
	\frac{\partial \mathcal I_1(\lambda)}{\partial x^{i}_{A}} &=&  p_{(2)\alpha}(\lambda) z^\alpha_{,(Ai)}(\lambda) + q^i_{(2)}(\lambda) - \Delta_r^{(1)}(\bm{z}(\lambda),t_b, \bm x_B) \left[ p_{(1)0\alpha}(\lambda) z^\alpha_{,(Ai)}(\lambda) + q^i_{(1)0} (\lambda) \right] \nonumber \\
	&&-  p_{(1)0}(\lambda) \; \dDdxa{i}\, ,\\
	\frac{\partial \mathcal I_1(\lambda)}{\partial x^{i}_{B}} &=&  p_{(2)\alpha}(\lambda) z^\alpha_{,(Bi)}(\lambda) - q^i_{(2)}(\lambda) - \Delta_r^{(1)}(\bm{z}(\lambda),t_b, \bm x_B) \left[ p_{(1)0\alpha}(\lambda) z^\alpha_{,(Bi)}(\lambda) - q^i_{(1)0} (\lambda) \right] \nonumber \\
	&&-  p_{(1)0}(\lambda) \; \dDdxb{i}\, 
\end{eqnarray}
and with $\Delta_r^{(1)}(\bz(\lambda),t_B,\bx_B)$ and its derivatives given by (\ref{eq:Deltaz}), (\ref{eq:dDz_dxia}) and (\ref{eq:dDz_dxib}). Similarly, we also compute
\begin{eqnarray}
\frac{\partial \mathcal I_2(\lambda)}{\partial x^{i}_{A}} &=& \left[ - N^i_{AB} \gab0j1 + \gab ij1 +(R_{AB} \dgab 0j1\alpha  - \dgab jk1\alpha R^k_{AB}) z^\alpha_{,(Ai)} \right]_{z^\beta(\lambda)}\; \times \dDdx{j}\nonumber\\
&& \quad + [ R_{AB} \gab 0j1 -  R^k_{AB} \gab jk1 ]_{z^\beta(\lambda)}\;\times \frac{\partial^2{\Delta}^{(1)}_{r}}{\partial x^{i}_{A}\partial x^{j}} (\bm z(\lambda),t_B,\bx_B) \,,\\
\frac{\partial \mathcal I_2(\lambda)}{\partial x^{i}_{B}} &=& \left[  N^i_{AB} \gab0j1 - \gab ij1 +(R_{AB} \dgab 0j1\alpha  - \dgab jk1\alpha R^k_{AB}) z^\alpha_{,(Bi)} \right]_{z^\beta(\lambda)}\; \times\dDdx{j}\nonumber\\
&& \quad + [ R_{AB} \gab 0j1 -  R^k_{AB} \gab jk1 ]_{z^\beta(\lambda)} \; \times\frac{\partial^2{\Delta}^{(1)}_{r}}{\partial x^{i}_{B}\partial x^{j}} (\bm z(\lambda),t_B,\bx_B) 
\end{eqnarray}
and %
\begin{eqnarray} 
 \frac{\partial \mathcal I_3 (\lambda)}{\partial x^{i}_{A}} &=&  \frac{N_{AB}^i}{2} \sum_{j=1}^3 {\left(\dDdx{j}\right)^2}- R_{AB} \sum_{j=1}^3 \left[ \dDdx{j}  \cdot \frac{\partial^2{\Delta}^{(1)}_{r}}{\partial x^{i}_{A}\partial x^{j}} (\bm z(\lambda),t_B,\bx_B ) \right] \, ,\\
 \frac{\partial \mathcal I_3 (\lambda)}{\partial x^{i}_{B}} &=& -\frac{N_{AB}^i}{2} \sum_{j=1}^3 {\left(\dDdx{j}\right)^2}- R_{AB} \sum_{j=1}^3 \left[ \dDdx{j}  \cdot \frac{\partial^2{\Delta}^{(1)}_{r}}{\partial x^{i}_{B}\partial x^{j}} (\bm z(\lambda),t_B,\bx_B ) \right] \, ,
\end{eqnarray}
where the derivatives of $\Delta_r^{(1)}$ are given by (\ref{eq:Dz_dx}) and the second derivatives will be given explicitly in Eq.~(\ref{eq:d2Delta}). Finally, the derivatives of $\mathcal I_j(\lambda)$ with respect to $t_B$ are given by
\begin{eqnarray}
\frac{\partial \mathcal I_1}{\partial t_B} &=&c\, p_{(2)0}(\lambda)  - c \, p_{(1)00} (\lambda) \,\Delta_r^{(1)}(z(\lambda),t_B,\bx_B) -   p_{(1)0}(\lambda)\, \frac{\partial \Delta_r^{(1)}}{\partial t_B}(\bm z(\lambda),t_B,\bx_B) \, , \\
	&& \nonumber\\ 
\frac{\partial \mathcal I_2}{\partial t_B} &=&c [ R_{AB} \dgab 0i10 - R^k_{AB} \dgab ik10]_{z^\beta(\lambda)}\times \dDdx i + [ R_{AB} \gab 0i1 - R^k_{AB} \gab ik1]_{z^\beta(\lambda)} \times \frac{\partial^2{\Delta}^{(1)}_{r}}{\partial t_{B}\partial x^{i}} (\bz(\lambda)) \, , \\
	&&\nonumber \\ 
\frac{\partial \mathcal I_3}{\partial t_B} &=& -R_{AB} \sum_{j=1}^3 {\dDdx{j} \cdot \frac{\partial^2{\Delta}^{(1)}_{r}}{\partial t_B\partial x^{j}} (\bz(\lambda),t_B,\bx_B )}  d\lambda \; ,
\end{eqnarray}
\end{subequations}          
where $\Delta^{(1)}_r$ is given by~(\ref{eq:Deltaz}), its first derivatives by~(\ref{eq:dDz_dtb}) and~(\ref{eq:Dz_dx}) and where the expression of the second derivatives are given by Eq.~(\ref{eq:d2Delta}).

The relations given above provide the TTF and its derivatives up to the 2PM order in an integral form particularly adapted for a numerically evaluation from any metric. When replaced into Eq.~\eqref{eq:doppler}, Eq.~\eqref{eq:astro} and Eq.~\eqref{eq:ang_sep}, they give a full description of Doppler and astrometric observables at 2PM. 


\section{Applications to a static, spherically symmetric space-time}\label{sec:applications}

\par The results presented above will be illustrated through the case of a static, spherically symmetric space-time. In isotropic coordinates, the line element can be written
\begin{equation}\label{eq:petric1}
	ds^2=A(r)c^2dt^2-B(r)\delta_{ij}dx^idx^j.
\end{equation}
As mentioned in \cite{linet:2013fk}, the light rays of metric (\ref{eq:petric1}) are the same as the light rays of any $d\tilde s^2$ conformal to (\ref{eq:petric1}). We can thus simplify the calculations by choosing $d\tilde s^2=A^{-1}(r)ds^2$ and deal with the following line element
\begin{equation}\label{eq:petric2}
	d\tilde s^2=c^2dt^2 - \frac{B(r)}{A(r)}\delta_{ij}dx^idx^j=c^2dt^2-U(r)\delta_{ij}dx^idx^j.
\end{equation}
We can now consider a PM expansion of the function $U(r)=1+U^{(1)}(r)+U^{(2)}(r)+\dots$. This procedure will simplify the results shown in Section~\ref{sec:der_ttf}. Let us assume that a light ray is emitted by ${\cal O}_{\cal A}$ at coordinates $(ct_A,{\bx}_A)$ and received by an observer ${\cal O}_{\cal B}$ at coordinates $(ct_B,{\bx}_B)$.  Using Eqs.~(\ref{eq:delta1})-(\ref{eq:dDelta1}), the reception delay function and its first derivatives at 1PM order can be written as
\begin{subequations}\label{eq:Delta1_SS}
	\begin{eqnarray}
		\Delta_r^{(1)}(\bx_A,\bx_B)&=&\frac{R_{AB}}{2}\int_0^1 U^{(1)}(z(\lambda))d\lambda \; , \label{eq:D1b}\\
		\frac{\partial \Delta_r^{(1)}}{\partial x^i_A}(\bx_A,\bx_B)&=&-\frac{U^{(1)}(r_A)}{2}N^i_{AB}+\left[\frac{R_{AB}}{2}x^i_B+\frac{N^i_{AB}}{4}(r_A^2-R^2_{AB}-r^2_B)\right]\times \int_0^1\frac{\lambda}{z(\lambda)}\frac{\partial U^{(1)}}{\partial r}(z(\lambda))d\lambda \; , \\
		\frac{\partial \Delta_r^{(1)}}{\partial x^i_B}(\bx_A,\bx_B)&=&\frac{U^{(1)}(r_A)}{2}N^i_{AB}+\frac{R_{AB}x^i_B}{2}\int_0^1\frac{1}{z(\lambda)}\frac{\partial U^{(1)}}{\partial r}(z(\lambda))d\lambda \\
		&&\qquad \qquad-\left[\frac{R_{AB}}{2}x^i_B+\frac{N^i_{AB}}{4}(r_A^2+R^2_{AB}-r^2_B)\right]\times \int_0^1\frac{\lambda}{z(\lambda)}\frac{\partial U^{(1)}}{\partial r}(z(\lambda))d\lambda \; , \nonumber
	\end{eqnarray}
\end{subequations}
where $z(\lambda)=\vert \bz(\lambda)\vert $, $\bz(\lambda)$ being given by Eq.~(\ref{eq:z}) and where we use the notations $r_A = \vert \bx_A \vert$, $r_B  = \vert \bx_B \vert$, $R_{AB}=\vert \bx_B-\bx_A \vert$ and $N_{AB}^i=(\bx_B^i-\bx_A^i)/R_{AB}$. Similarly, using (\ref{eq:Dz_dx}), one can show that
\begin{equation}\label{eq:dDdx2_s}
 \frac{\partial \Delta_r^{(1)}}{\partial x^i}(\bz(\lambda),\bx_B)=-U^{(1)}(z(\lambda))\frac{N^i_{AB}}{2}+\lambda \left[\frac{R_{AB}}{2}x^i_B + \frac{N^i_{AB}}{4}(r_A^2-R_{AB}^2-r_B^2)\right]V(\lambda) \, ,
\end{equation}
with
\begin{equation}\label{eq:V}
	V(\lambda) \equiv \int_0^1\frac{\mu}{z(\lambda\mu)}\frac{\partial U^{(1)}}{\partial r}(z(\lambda\mu))d\mu \; .
\end{equation}
Substituting now for the metric tensor from Eq.~\eqref{eq:petric2} into Eq.~(\ref{eq:delta2}), the 2PM order of the reception delay function is given by 
\begin{equation}\label{eq:Delta2_SS}
	\Delta_r^{(2)}(\bx_A,\bx_B)=\frac{R_{AB}}{2}\int_0^1 \left[ U^{(2)}(z(\lambda))+\bar {\mathcal I}_3(\lambda)\right]d\lambda \, ,
\end{equation}
where we defined $\bar {\mathcal I}_3(\lambda) \equiv 2 {\mathcal I}_3(\lambda)/ R_{AB}$. Using Eq.~(\ref{eq:dDdx2_s}), one gets
\begin{eqnarray}
\bar{	\mathcal I}_3(\lambda)&=& -\sum_{j=1}^3\left[\frac{\partial \Delta_r^{(1)}(\bz(\lambda))}{\partial x^j}\right]^2\nonumber\\
&=&-\frac{1}{4}\left\{ U_{(1)}^2(z(\lambda)) + \frac{V^2(\lambda)}{4}\Big[4r_B^2R_{zB}^2-(z^2(\lambda)-R^2_{zB}-r_B^2)^2\Big]\right\}\label{eq:I3}
\end{eqnarray}
with  $R_{zB} \equiv \vert\bx_B-\bz(\lambda)\vert$. In the last relation, it can sometimes be useful to replace $z(\lambda)^2-R^2_{zB}-r_B^2=\lambda(r_A^2-R_{AB}^2-r_B^2)$ or $4R^2_{zB}r_B^2-(z^2(\lambda)-R^2_{zB}-r_B^2)^2=-\lambda^2\left[(r_A+r_B)^2-R_{AB}^2\right]\left[(r_A-r_B)^2-R_{AB}^2\right]$ and use $V(\lambda)$ as defined by (\ref{eq:V}). From Eqs.~\eqref{dDelta2}-\eqref{k2PM}, the derivatives of $\Delta_r^{(2)}$ are then given by
\begin{subequations}
	\begin{eqnarray}
		\frac{\partial \Delta_r^{(2)}}{\partial x^i_A} &=&-\frac{N^i_{AB}}{R_{AB}}\Delta_r^{(2)}+\frac{R_{AB}}{2}\int_0^1\left[\frac{\lambda z^i(\lambda)}{z(\lambda)}\frac{\partial U^{(2)}}{\partial r}(z(\lambda))+\frac{\partial \bar{\mathcal I}_3}{\partial x^i_A}(\lambda)\right] d\lambda\label{eq:dD2_dxia} \, ,\\
				\frac{\partial \Delta_r^{(2)}}{\partial x^i_B} &=&\frac{N^i_{AB}}{R_{AB}}\Delta_r^{(2)}+\frac{R_{AB}}{2}\int_0^1\left[\frac{(1-\lambda) z^i(\lambda)}{z(\lambda)}\frac{\partial U^{(2)}}{\partial r}(z(\lambda))+\frac{\partial \bar{\mathcal I}_3}{\partial x^i_B}(\lambda)\right] d\lambda \; ,
	\end{eqnarray}
\end{subequations}
with
\begin{subequations}
	\begin{eqnarray}
		\frac{\partial \bar{\mathcal I}_3}{\partial x^i_A}(\lambda)&=&-\frac{1}{4} \bigg\{ 2\lambda\frac{z^i(\lambda)}{z(\lambda)}U^{(1)}(z(\lambda))\frac{\partial U^{(1)}}{\partial r}(z(\lambda)) -\lambda^2\frac{V(\lambda)}{2}\frac{\partial V}{\partial x^i_A}(\lambda)\left[(r_A+r_B)^2-R_{AB}^2\right]\left[(r_A-r_B)^2-R_{AB}^2\right] \nonumber \\ 
		&&  \qquad \qquad \qquad -\lambda^2 V^2(\lambda)\left[2r_B^2 R^i_{AB}+(r_A^2-R^2_{AB}-r^2_B)x^i_B\right]\bigg\} \, , \label{eq:di3dxia} \\
				\frac{\partial \bar{\mathcal I}_3}{\partial x^i_B}(\lambda)&=&-\frac{1}{4} \bigg\{ 2(1-\lambda)\frac{z^i(\lambda)}{z(\lambda)}U^{(1)}(z(\lambda))\frac{\partial U^{(1)}}{\partial r}(z(\lambda)) -\lambda^2\frac{V(\lambda)}{2}\frac{\partial V}{\partial x^i_B}(\lambda)\left[(r_A+r_B)^2-R_{AB}^2\right]\left[(r_A-r_B)^2-R_{AB}^2\right] \nonumber \\ 
				&&  \qquad \qquad \qquad +\lambda^2 V^2(\lambda)\left[ (r_A^2+r_B^2-R^2_{AB}) R^i_{AB}+(r_A^2+R^2_{AB}-r^2_B)x^i_B\right]\bigg\} \,.
	\end{eqnarray}
\end{subequations}
and where the derivatives of $V(\lambda)$ can be computed as
\begin{subequations}\label{eq:dV}
	\begin{eqnarray}
		\frac{\partial V}{\partial x^i_A}(\lambda)&=&\int_0^1 \left[\frac{\partial^2 U^{(1)}}{\partial r^2}(z(\lambda\mu))\frac{\lambda\mu^2z^i(\lambda\mu)}{z^2(\lambda\mu)}-\lambda\mu^2\frac{\partial U^{(1)}}{\partial r}(z(\lambda\mu))\frac{z^i(\lambda\mu)}{z^3(\lambda\mu)}\right]d\mu \, ,\\
		\frac{\partial V}{\partial x^i_B}(\lambda)&=&\int_0^1 \left[\frac{\partial^2 U^{(1)}}{\partial r^2}(z(\lambda\mu))\frac{(1-\lambda\mu)\mu z^i(\lambda\mu)}{z^2(\lambda\mu)}-(1-\lambda\mu)\mu\frac{\partial U^{(1)}}{\partial r}(z(\lambda\mu))\frac{z^i(\lambda\mu)}{z^3(\lambda\mu)}\right]d\mu \, .
	\end{eqnarray}
\end{subequations}

Let us now study a Schwarzschild-like metric, whose expansion in isotropic coordinates is
\begin{equation}\label{eq:schwarz}
	ds^2=\left(1-2\frac{m}{r}+2\beta\frac{m^2}{r^2}+\dots\right) c^2 dt^2 -\left(1+2\gamma\frac{m}{r}+\frac{3}{2}\epsilon\frac{m^2}{r^2}+\dots\right)\delta_{ij}dx^idx^j \; ,
\end{equation}
and $U(r)$ is given by
	\begin{equation}\label{eq:U_S}
	U(r)=1+2(1+\gamma)\frac{m}{r}+2\kappa\frac{m^2}{r^2}+\dots \, ,
\end{equation}
where $\kappa=2(1+\gamma)-\beta+\frac{3}{4}\epsilon$.

Introducing $U(r)$ from Eq.~(\ref{eq:U_S}) into (\ref{eq:Delta1_SS}) leads to
\begin{subequations}\label{eq:D1_s}
	\begin{eqnarray}
		\Delta_r^{(1)}&=&R_{AB}(1+\gamma)m \int_0^1 \frac{d\lambda}{z(\lambda)}=(\gamma+1)m\ln \left( \frac{r_A+r_B+R_{AB}}{r_A+r_B-R_{AB}}\right) \, ,\label{eq:D1_stof}\\
		\frac{\partial \Delta_r^{(1)}}{\partial x^i_A}&=&-(1+\gamma)\frac{m}{r_A}N^i_{AB}+\left[\frac{R_{AB}}{2}x^i_B+\frac{N^i_{AB}}{4}(r_A^2-R^2_{AB}-r^2_B)\right]\times\frac{-4(1+\gamma)m}{r_A\left[(r_A+r_B)^2-R_{AB}^2\right]} \nonumber\\
		&=&-\frac{2(1+\gamma)m}{(r_A+r_B)^2-R_{AB}^2}\left[\frac{R_{AB}}{r_A}x^i_A + N^i_{AB}(r_A+r_B) \right] \label{eq:dDdxb1tof}\, ,\\
 \frac{\partial \Delta_r^{(1)}}{\partial x^i_B}&=&(1+\gamma)\frac{m}{r_A}N^i_{AB}-\frac{R_{AB}x^i_B}{2}\left[\frac{4(1+\gamma)m\left(\frac{1}{r_A}+\frac{1}{r_B}\right)}{(r_A+r_B)^2-R_{AB^2}}\right] \nonumber\\
&&\qquad -\left[\frac{R_{AB}}{2}x^i_B+\frac{N^i_{AB}}{4}(r_A^2+R^2_{AB}-r^2_B)\right]\times\frac{-4(1+\gamma)m}{r_A\left[(r_A+r_B)^2-R_{AB}^2\right]} \nonumber\\
		&=&-\frac{2(1+\gamma)m}{(r_A+r_B)^2-R_{AB}^2}\left[\frac{R_{AB}}{r_B}x^i_B - N^i_{AB}(r_A+r_B) \right] \, \label{eq:dDdxb1}
	\end{eqnarray}
\end{subequations}
One should note that Eq.~(\ref{eq:D1_stof}) is equivalent to the expression of the time delay found by Shapiro \cite{shapiro:1964kh}, while the two derivatives~(\ref{eq:dDdxb1tof}) and (\ref{eq:dDdxb1}) are in agreement with results found in \cite{blanchet:2001ud}. 

The computation at the 2PM order is more cumbersome. Substituting for $U(r)$ from Eq.~(\ref{eq:U_S}) into Eq.~(\ref{eq:Delta2_SS}), one gets
\begin{equation}\label{eq:SD2_1}
	\Delta_r^{(2)}=R_{AB}\kappa \, m^2\int_0^2\frac{d\lambda}{z^2(\lambda)}+\frac{R_{AB}}{2}\int_0^1 \bar{\mathcal I}_3(\lambda)d\lambda
\end{equation}
where, using $V(\lambda)$ as determined from Eq.~(\ref{eq:V})
\begin{equation}
	V(\lambda)=-2(1+\gamma)m\int_0^1\frac{\mu}{z^3(\lambda\mu)}d\mu=-\frac{4(1+\gamma)m}{z(\lambda)\left[(z(\lambda)+r_B)^2-\lambda^2R_{AB}^2\right]}.
\end{equation}
into Eq.~\eqref{eq:I3}, we obtain $\bar{\mathcal I}_3(\lambda)$ as
\begin{equation}
\bar{	\mathcal I}_3(\lambda)=-\frac{4(1+\gamma)^2m^2r_B}{z(\lambda)\left[(z(\lambda)+r_B)^2-\lambda^2R_{AB}^2\right]}=-4(1+\gamma)^2m^2\frac{d}{d\lambda}\left[\frac{\lambda}{(z(\lambda)+r_B)^2-\lambda^2R_{AB}^2}\right] \; .
\end{equation}
Replacing this expression in Eq.~(\ref{eq:SD2_1}) and integrating, one gets
\bea 
&&\Delta_r^{(2)}(\bx_A, \bx_B) = m^2\frac{R_{AB}}{r_{A} r_{B}}\left[
\frac{\kappa\arccos \mu}
{\sqrt{1-\mu^2}}
-\frac{(1+\gamma)^2}{1 + \mu}\right],  \label{eq:D2b} \\
&&\nonumber
\eea
with $\mu=(\bn_A . \bn_B)$ and where $\bn_{A/B}=\bx_{A/B}/r_{A/B}$. Substituting for $\Delta_r$ from Eqs~\eqref{eq:D1b}-\eqref{eq:D2b} into Eq.~\eqref{eq:TTF_PM}, we finally get an expression for the TTF in a Schwarzschild-like metric and up to 2PM as
\begin{equation}\label{eq:range_S2}
	\mathcal T_r(\bx_A,t_B,\bx_B)=t_B-t_A=\frac{R_{AB}}{c}+\frac{(\gamma+1)m}{c}\ln \left( \frac{r_A+r_B+R_{AB}}{r_A+r_B-R_{AB}}\right)+\frac{m^2R_{AB}}{c\, r_Ar_B}\left[
	\frac{\kappa\arccos \mu}
	{\sqrt{1-\mu^2}}
	-\frac{(1+\gamma)^2}{1 + \mu}\right].
\end{equation}
We recover a result previously derived by different approaches \cite{richter:1983oq,le-poncin-lafitte:2004cr,teyssandier:2008nx,ashby:2010fk,linet:2013fk} (see also  \cite{brumberg:1987bh} in the case where $\beta = \gamma = \delta = 1$).
\bigskip

We can now compute the derivatives of $\Delta_r^{(2)}$. As an example, we will only focus on the derivative with respect to $x^i_A$; the other derivative (with respect to $x^i_B$) can be computed similarly. Using Eq.~\eqref{eq:U_S} into Eq.~(\ref{eq:dV}), one gets
\begin{equation}
	\frac{\partial V}{\partial x^i_A}(\lambda)=\frac{8(1+\gamma)m\lambda}{z^3(\lambda)\left[(z(\lambda)+r_B)^2-\lambda^2R_{AB}^2\right]^2}\Big\{\left[(z^2(\lambda)+2z(\lambda)r_B+\bz(\lambda)\cdot\bx_B\right]x^i_B-\lambda R^i_{AB}\left[2z(\lambda)r_B+\bz(\lambda)\cdot\bx_B\right]\Big\}.
\end{equation}
Replacing this result in Eq.~(\ref{eq:di3dxia}) then leads to
\begin{equation}
	\frac{\partial \bar{\mathcal I}_3}{\partial x^i_A}= \frac{4(1+\gamma)^2m^2r_B\lambda}{z^3(\lambda)\Big[(z(\lambda)+r_B)^2-\lambda^2R_{AB}^2\Big]^2}\Bigg\{x^i_B\Big[ r^2_B+4r_Bz(\lambda)+3z^2(\lambda)-\lambda^2R_{AB}^2\Big] -\lambda R^i_{AB}\Big[ r^2_B+4r_Bz(\lambda)+z^2(\lambda)-\lambda^2R_{AB}^2\Big] \Bigg\} \; ,
\end{equation}
which, after some lengthy but straightforward calculations, can be written as
\begin{equation}\label{eq:di3_s}
	\frac{\partial \bar{\mathcal I}_3}{\partial x^i_A}= 8(1+\gamma)^2m^2\frac{d}{d\lambda}\left[\lambda^2\frac{(z(\lambda)+r_B)x^i_B-\lambda r_BR^i_{AB}}{z(\lambda)\Big[(z(\lambda)+r_B)^2-\lambda^2R_{AB}^2\Big]^2}\right] \; .
\end{equation}
Finally, one needs to compute the integral corresponding to the second term of Eq.~(\ref{eq:dD2_dxia}), namely
\begin{eqnarray}
&& \!\!\!\!\!\!\!\!\!\!\!\!\!\!\!\!\!\!\!\!\!\!\!\!\!\!\!\!\!\!\!\!\!\!\!\!\!\!\!\!  \frac{R_{AB}}{2} \int_0^1\left[\frac{\lambda z^i(\lambda)}{z(\lambda)}\frac{\partial U^{(2)}}{\partial r}(z(\lambda))\right]d\lambda	=-2\kappa \,  R_{AB} m^2\int_0^1\frac{\lambda z^i(\lambda)}{z^4(\lambda)}d\lambda\nonumber\\
&& \qquad\qquad\qquad\qquad\; =\frac{\kappa \, m^2 R_{AB} \, \arccos \mu}{r^2_Ar_B(1-\mu^2)^{3/2}}\left(-n^i_A+\mu n^i_B\right)-\frac{\kappa \, m^2 R_{AB}}{r_A^2r_B(1-\mu^2)}(n^i_B-\mu n^i_A).\label{eq:dint_s}
\end{eqnarray}
Now, substituting from Eq.~(\ref{eq:D2b}), Eq.~(\ref{eq:di3_s}) and Eq.~(\ref{eq:dint_s}) into Eq.~(\ref{eq:dD2_dxia}), one gets
\begin{subequations}\label{eq:D2_s}
\begin{eqnarray}
	\frac{\partial \Delta_r^{(2)}}{\partial x^i_A}&=&\frac{\kappa \, m^2}{r_Ar_B}\left\{\frac{\arccos \mu}{\sqrt{1-\mu^2}}\left[-N^i_{AB}-\frac{R_{AB}}{r_A(1-\mu^2)}\left(n^i_A-\mu n^i_B\right)\right]-\frac{R_{AB}}{r_A(1-\mu^2)}(n^i_B-\mu n^i_A)\right\}\nonumber\\
&&+\frac{(1+\gamma)^2m^2}{r_Ar_B(1+\mu)}\left\{N^i_{AB}+\frac{R_{AB}}{r_A(1+\mu)}(n^i_A+n^i_B)\right\} \, ,
\end{eqnarray}
while a similar reasoning for $\frac{\partial \Delta_r^{(2)}}{\partial x^i_B}$ lead to 
\begin{eqnarray}
	\frac{\partial \Delta_r^{(2)}}{\partial x^i_B}&=&\frac{\kappa \, m^2}{r_Ar_B}\left\{\frac{\arccos \mu}{\sqrt{1-\mu^2}}\left[ N^i_{AB}-\frac{R_{AB}}{r_B(1-\mu^2)} \left(n^i_B-\mu n^i_A\right) \right] -\frac{R_{AB}}{r_B(1-\mu^2)}(n^i_A-\mu n^i_B)\right\}\nonumber\\
&&+\frac{(1+\gamma)^2m^2}{r_Ar_B(1+\mu)}\left\{-N^i_{AB}+\frac{R_{AB}}{r_B(1+\mu)}(n^i_A+n^i_B)\right\}. \label{eq:dDdxb2}
\end{eqnarray}
\end{subequations}
Some algebra allows to put the last two results in the same form as the one found in \citep{teyssandier:2012uq}, which serves as verification of our approach. Of course, in the case of the Schwarzschild metric the analytical derivation of Eq.~(\ref{eq:D2b}) is much simpler than the above calculations to get Eq.~\eqref{eq:D2_s} and can be used to check our calculation. Nevertheless the method presented here is very efficient for numerical evaluations of the derivatives of the TTF, necessary when using more complex metrics and for the test of alternative theories of gravity, when the integrals are no longer analytic. 
As an example, we will present in this section several applications of our formulae to future space missions.

\subsection{Application to BepiColombo}
The future BepiColombo mission will reach an impressive level of accuracy on its measurements: $10 \,\rm{cm}$ on the range and $10^{-6}\, \rm{m/s}$ on the Doppler~\citep{milani:2002vn,iess:2009fk}. Such an accuracy needs a light propagation model that includes the influence of some of the 2PM terms coming from the Sun~\citep{tommei:2010uq}. As an example of how the equations presented in this paper can be applied to a real measurement, we simulate a one year Mercury-Earth Doppler link taking into account only the gravitational contribution from the Sun. The Earth and Mercury orbits used here come from the JPL ephemerides \citep{folkner:2009fk,folkner:2010kx} obtained using the SPICE toolkit \citep{acton:1996fk}. 

\begin{figure}[hbt]
\begin{center}
\includegraphics[width=0.45\textwidth]{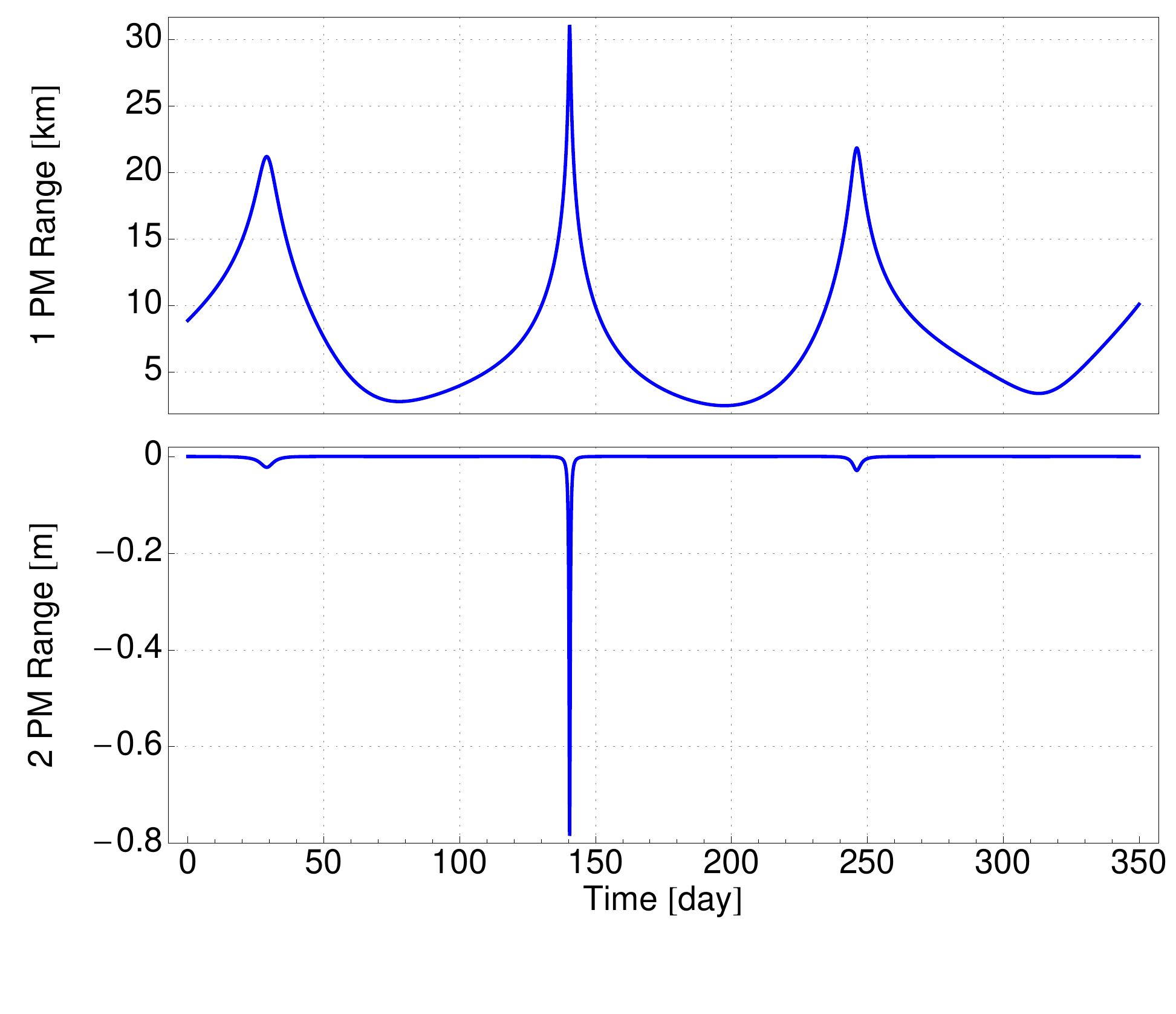}\hfill
\includegraphics[width=0.45\textwidth]{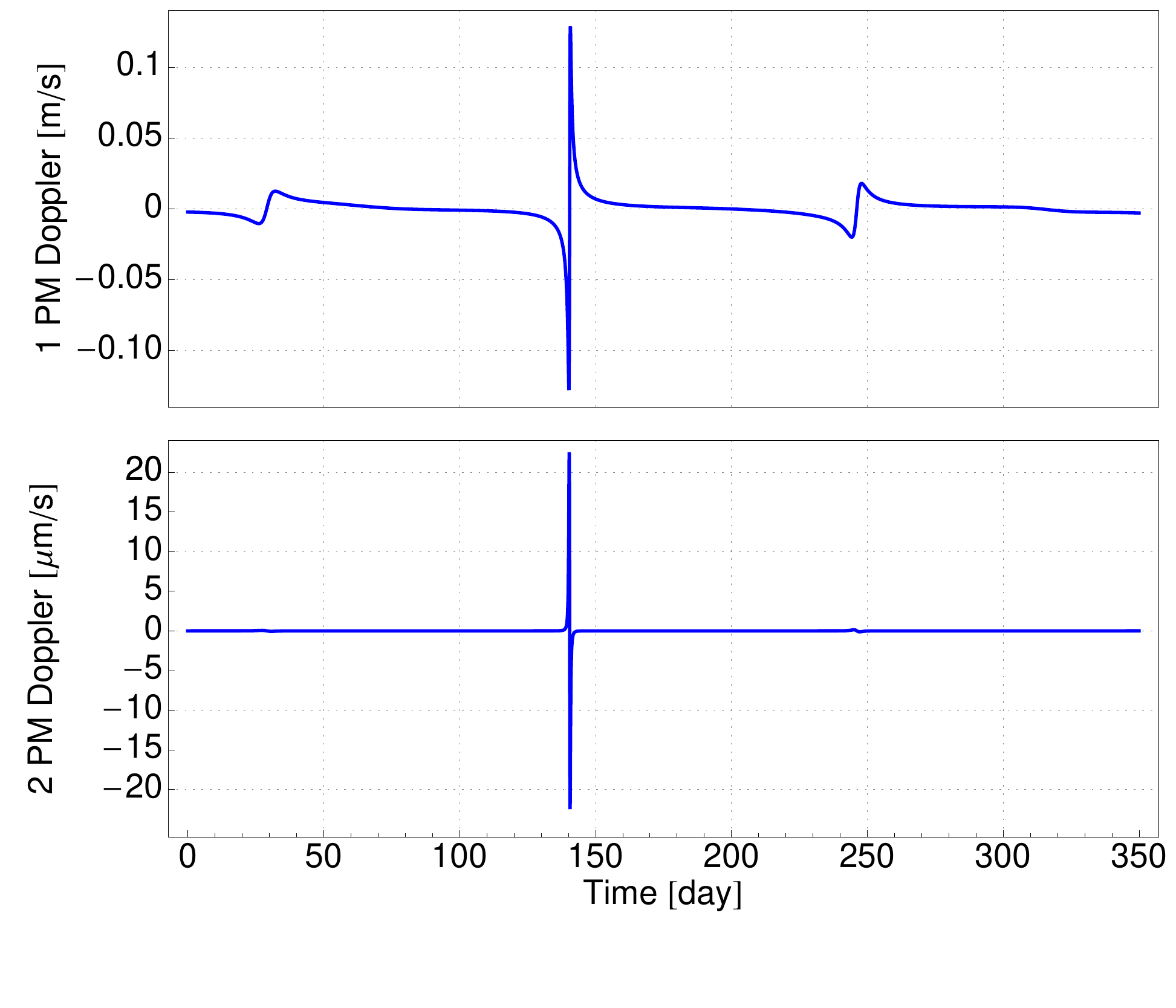}
\end{center}
\caption{First and second post Minkowskian contributions to the Range and the Doppler for a 1 year Mercury-Earth radioscience link.}
\label{fig:RD}
\end{figure}

\begin{figure}[htb]
\begin{center}
\includegraphics[width=0.6\textwidth]{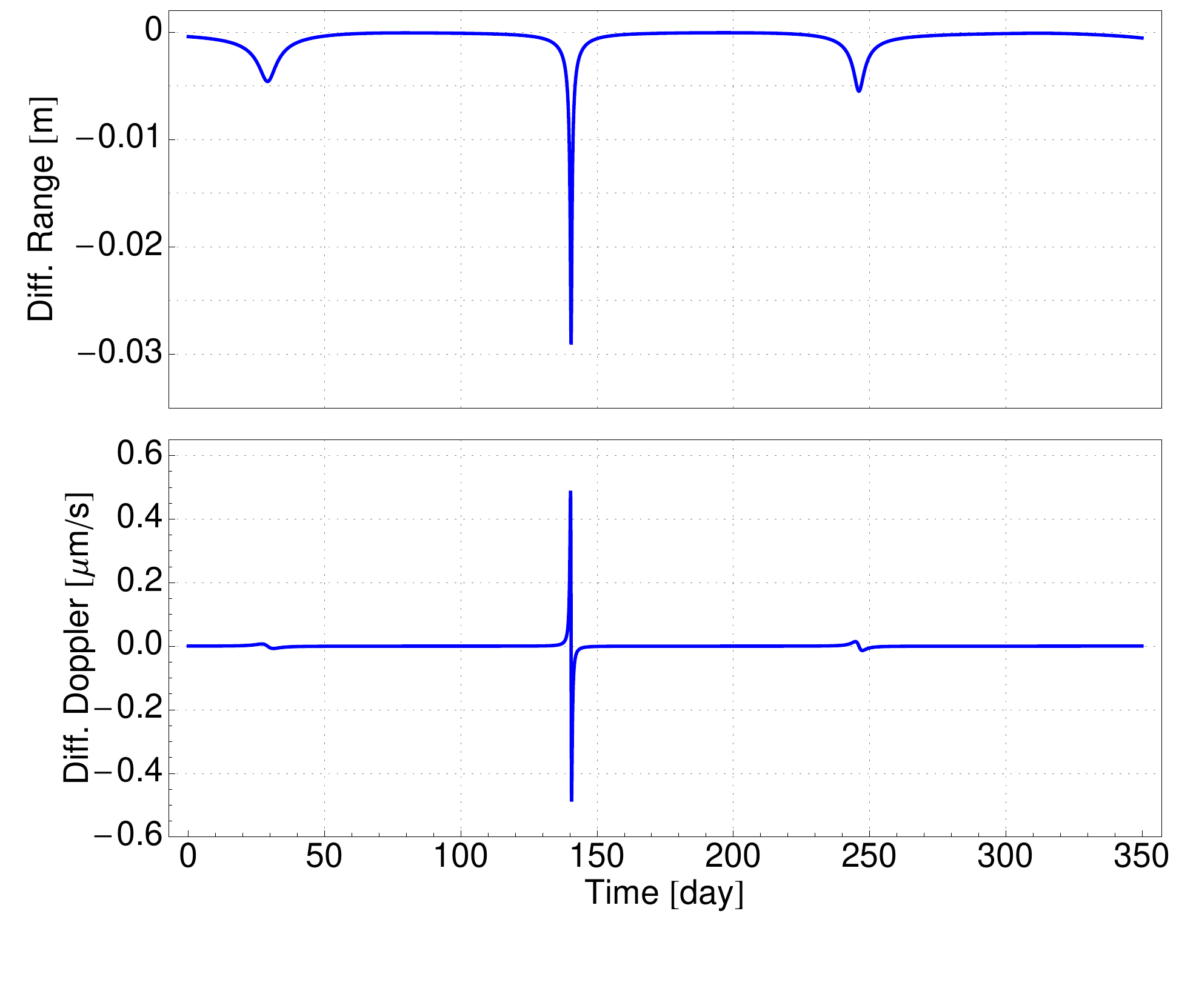}
\end{center}
\caption{Difference between the standard formulation of the Range/Doppler used in radioscience modeling (\ref{eq:range_Moyer}) and the exact 2PM expression (\ref{eq:range_S2}).} 
\label{fig:RDdiff}
\end{figure}

Substituting for the metric, $\Delta_r$ and its derivatives from Eq.~(\ref{eq:schwarz}), Eq.~(\ref{eq:D1_s}) and Eq.~(\ref{eq:D2_s}), respectively into Eq.~(\ref{eq:doppler}) one can write the expression of the Doppler around a spherical mass as
\begin{eqnarray}\label{eq:dopp_S_2pm}
	\frac{\nu_B}{\nu_A}&=&\frac{\sqrt{1-2\dfrac{m}{r_A}+2\beta\dfrac{m^2}{r^2_A}-\dfrac{3}{2}\beta_3\dfrac{m^3}{r^3_A}-\dfrac{v^2_A}{c^2}-2\gamma\dfrac{v_A^2}{c^2}\dfrac{m}{r_A}-\dfrac{3}{2}\epsilon\dfrac{m^2}{r_A^2}\dfrac{v_A^2}{c^2}}}{\sqrt{1-2\dfrac{m}{r_B}+2\beta\dfrac{m^2}{r^2_B}-\dfrac{3}{2}\beta_3\dfrac{m^3}{r^3_B}-\dfrac{v_B^2}{c^2}-2\gamma\dfrac{v_B^2}{c^2}\dfrac{m}{r_B}-\dfrac{3}{2}\epsilon\dfrac{m^2}{r_B^2}\dfrac{v_B^2}{c^2}}}  \times\dfrac{q_B}{q_A} \, ,
\end{eqnarray} 
where we defined
\begin{subequations}\label{eq:dopp_S_2pm_2}
	\begin{eqnarray}
		q_A&=& 1- \frac{\bN_{AB}\cdot \bv_A}{c}-\frac{(1+\gamma)m}{cr_Ar_B(1+\mu)}\Big[(r_A+r_B)\bN_{AB}\cdot\bv_A+R_{AB}\bn_A\cdot \bv_A\Big] \nonumber \\
		&&+\frac{\kappa m^2}{c\, r_Ar_B}\left[\frac{\arccos \mu}{\sqrt{1-\mu^2}}\left(-\bN_{AB}\cdot\bv_A-\frac{R_{AB}}{r_A(1-\mu^2)}\left(\bn_A\cdot\bv_A - \mu \bn_B\cdot\bv_A\right)\right)-\frac{R_{AB}}{r_A(1-\mu^2)}(\bn_B\cdot\bv_A-\mu \bn_A\cdot\bv_A)\right] \nonumber \\
		&&+\frac{(1+\gamma)^2m^2}{c\, r_Ar_B(1+\mu)}\left[\bN_{AB}\cdot\bv_A+\frac{R_{AB}}{r_A(1+\mu)}\left(\bn_A\cdot\bv_A+\bn_B\cdot\bv_A\right)\right] 
	\end{eqnarray}
	and
	\begin{eqnarray}
		q_B&=&1-\frac{\bN_{AB}\cdot \bv_B}{c}-\frac{(1+\gamma)m}{cr_Ar_B(1+\mu)}\left[(r_A+r_B)\bN_{AB}\cdot\bv_B-R_{AB}\bn_B\cdot\bv_B \right] \nonumber \\
&&+\frac{\kappa m^2}{c\, r_Ar_B}\left[\frac{\arccos \mu}{\sqrt{1-\mu^2}}\left(-\bN_{AB}\cdot\bv_B+\frac{R_{AB}}{r_B(1-\mu^2)}\left(\bn_B\cdot\bv_B-\mu \bn_A\cdot\bv_B\right)\right)+\frac{R_{AB}}{r_B(1-\mu^2)}(\bn_A\cdot\bv_B-\mu \bn_B\cdot\bv_B)\right] \nonumber \\
		&&+\frac{(1+\gamma)^2m^2}{c\, r_Ar_B(1+\mu)}\left[\bN_{AB}\cdot\bv_B-\frac{R_{AB}}{r_B(1+\mu)}\left(\bn_A\cdot\bv_B+\bn_B\cdot\bv_B\right)\right] \, .
	\end{eqnarray}
\end{subequations}
We use relation (\ref{eq:range_S2}) and Eq.~\eqref{eq:dopp_S_2pm}-\eqref{eq:dopp_S_2pm_2} to estimate the order of magnitude of the first and second PM contributions to the Mercury-Earth range and Doppler as illustrated in Figure~\ref{fig:RD}. The different peaks correspond to Solar conjunctions in the geometry of the observation. 
Moreover, we would like to stress the fact that the expression of the time transfer used in the standard modeling of radioscience measurements (see for example \citep{moyer:2000uq}) is only an approximation of the relation (\ref{eq:range_S2}) given by
\begin{equation}\label{eq:range_Moyer}
	\mathcal T_r(\bx_A,t_B,\bx_B)=t_B-t_A=\frac{R_{AB}}{c}+\frac{(\gamma+1)m}{c}\ln \left( \frac{r_A+r_B+R_{AB}+(1+\gamma)m}{r_A+r_B-R_{AB}+(1+\gamma)m}\right).	
\end{equation}
A comparison of range and Doppler simulations obtained using expressions based on the approximation~(\ref{eq:range_Moyer}) and on expression~(\ref{eq:range_S2}), which is complete up to 2PM order, is shown in Figure~\ref{fig:RDdiff} to quantify the accuracy of the standard radioscience modeling. We get results just below BepiColombo accuracy. Nevertheless, future space missions are going to aim at increasing the level of accuracy on radioscience measurements so that the current modeling shall be improved to include the full 2PM correction on light propagation.

\subsection{Direction of a light ray emitted by a star and observed on Earth}

In order to simulate an astrometric observable, one can specify the reference frame used to give the incident direction of a light ray. As shown in section~\ref{sec:astroobs}, this reference frame is mathematically modeled by a tetrad $E^\mu_{\left< \alpha \right>}$, which explicitly appears in the computation of the astrometric observables (\ref{eq:astro}). We develop here the expression of a kinematically nonrotating tetrad comoving with an observer in the case of a static spherically symmetric space-time described by the metric (\ref{eq:petric1}). This tetrad is called "kinematically nonrotating" in the sense that the spatial coordinates transformation between the global and the local coordinate frames does not depend on a time dependent orthogonal matrix \citep{klioner:1998zl}. This kind of local coordinate system is currently used in the definition of the Celestial Geocentric Reference System~\citep{soffel:2003bd} and is extensively used in the context of the Gaia mission~\citep{klioner:2004fk}. Defining $\partial _\alpha$ the vectors of the natural coordinate basis and $e_{\left< \alpha \right>}$ the basis vectors of the tetrad, the transformation between these two basis is noted $E_{\left< \alpha \right>}^\mu$ and is given by 
\be \lb{eq:bastransmatrix}
	e_{\left< \alpha \right>} = E_{\left< \alpha \right>}^\mu \partial _\mu \; .
\ee
The great advantage of such a basis is that the tetrad is locally orthonormal. This transformation physically corresponds to a change of basis in the tangent space of the differential manifold.
From the point of view of the metric, we can easily show the link between the $g_{\mu \nu}$ of the natural coordinate basis and $\eta_{\left< \alpha \right>\left< \beta \right>}$ using Eq.~(\ref{eq:bastransmatrix})
\be
	\eta_{\left< \alpha \right>\left< \beta \right>} = \bm g (\bm e_{\left< \alpha \right>},\bm e_{\left< \beta \right>} ) =  \bm g (E_{\left< \alpha \right>}^\mu \partial _\mu,E_{\left< \beta \right>}^\nu \partial _\nu) =  E_{\left< \alpha \right>}^\mu E_{\left< \beta \right>}^\nu \bm g ( \partial _\mu, \partial _\nu) =  E_{\left< \alpha \right>}^\mu E_{\left< \beta \right>}^\nu  g_{\mu \nu} \, .
\ee
All indexes related to the tetrad (between angle brackets) are raised and lowered using Minkowsky metric tensor, while natural coordinate basis indexes are set up and down using the $g_{\mu \nu}$ metric. 

We can split the transformation between the natural coordinate basis and the local comoving basis of the tetrad into two parts $\lambda^{\mu}_{\left< \alpha \right>} =\Lambda^{\hat \kappa}_{\left< \alpha \right>} \tilde \Lambda^\mu_{\hat \kappa}$ \citep{misner:1973fk}. The first step (parametrized by $ \tilde\Lambda^\mu_{\hat \kappa}$) consists in orthogonalizing the natural coordinate basis to obtain a local orthonormal coordinate basis static with respect to the coordinate system used. The second part of the transformation (parametrized by $\Lambda^{\hat \kappa}_{\left< \alpha \right>}$) consists in applying a Lorentz boost to this orthonormal basis to make it comoving with the observer. Quantities related to the final tetrad will be denoted with indices between angle brackets while quantities expressed in the intermediate tetrad will be denoted with a hat. Since the space-time metric (\ref{eq:petric1}) is diagonal, it is straightforward to orthonormalize the basis
\bea \label{eq:tetrad1}
	\tilde\Lambda_{\hat 0}^0 = \frac{1}{\sqrt{A(r)}} \qquad , \qquad	\tilde\Lambda_{\hat 0}^i = \tilde\Lambda_{\hat i}^0 = 0 \qquad , \qquad	\tilde\Lambda_{\hat i}^j = \frac{\delta_{\hat i}^j }{\sqrt{B(r)}}. 
\eea
The second step consists in  a Lorentz boost of the previous tetrad in order to make it comoving with the observer. We will note the quadri-velocity of the observer (expressed in the global coordinate system) by $u^\alpha=dx^\alpha/ds$. This velocity can also be expressed in terms of coordinates related to the intermediate tetrad $\hat u^{\hat\alpha}=d\hat x^{\hat\alpha}/ds=\tilde\Lambda^{\hat \alpha}_\mu u^\mu=\left(\sqrt{A(r)}u^0,\sqrt{B(r)}u^i\right)$. Finally, the coordinate velocity of the observer will be denoted by $\beta^i=\dfrac{1}{c}\dfrac{dx^i}{dt}$. The same quantity expressed in the intermediate tetrad is $\hat \beta^i=\dfrac{1}{c}\dfrac{d\hat x^i}{d\hat t}=\sqrt{\dfrac{B(r)}{A(r)}}\beta^i$. The second matrix transformation is thus simply given by a standard Lorentz transformation matrix whose inverse is given by
\bea \label{eq:tetrad2}
	\Lambda^{\hat 0}_{\left< 0 \right>} = \hat \gamma \qquad , \qquad  \Lambda^{\hat 0}_{\left< i \right>} = \Lambda^{\hat i}_{\left< 0 \right>} = -\hat \gamma  \hat \beta^i  \qquad , \qquad	\Lambda^{\hat i}_{\left< j \right>} = \delta_{ij} + \frac{\hat \gamma^2}{\hat \gamma + 1} \hat \beta^i \hat \beta^j 
\eea
with
\bea
	\hat \gamma &=& \Big( 1-\hat \beta^2 \Big)^{-1/2} = \Bigg(1-\frac{B(r)}{A(r)} \beta^2 \Bigg)^{-1/2} \; .
\eea
The combination of Eq.~(\ref{eq:tetrad1}) and Eq.~(\ref{eq:tetrad2}) gives
\begin{subequations}\label{eq:tetrad}
	\begin{eqnarray}
		E^0_{\left< 0 \right>}&=&\frac{\hat\gamma}{\sqrt{A(r)}}=\frac{1}{\sqrt{A(r)-B(r)\beta^2}} \; , \\
		E^i_{\left< 0 \right>}&=&-\frac{\hat \gamma  \hat \beta^i }{\sqrt{B(r)}}= -\frac{\beta^i}{\sqrt{A(r)-B(r)\beta^2}}  \; , \\
		E^0_{\left< j \right>}&=&-\frac{\hat \gamma  \hat \beta^j }{\sqrt{A(r)}}= -\sqrt{\frac{B(r)}{A(r)}}\frac{\beta^j}{\sqrt{A(r)-B(r)\beta^2}}  \; , \\
		E^i_{\left< j \right>}&=& \frac{\delta_{ij} + \dfrac{\hat \gamma^2}{\hat \gamma + 1} \hat \beta^i \hat \beta^j}{\sqrt{B(r)}} = \frac{\delta_{ij}}{\sqrt{B(r)}} + \frac{\sqrt{B(r)}\beta^i\beta^j}{\sqrt{A^2(r)-A(r)B(r)\beta^2}+A(r)-B(r)\beta^2} \; .
\end{eqnarray}
\end{subequations}
Eq.~\eqref{eq:tetrad} is the exact expression of a kinematically nonrotating tetrad comoving with a given observer in a static, spherically symmetric space-time. It can be expanded to 2PM order if necessary using Eqs.~\eqref{eq:schwarz}-\eqref{eq:U_S}.

\begin{figure}[hbt]
\begin{center}
\includegraphics[width=0.45\textwidth]{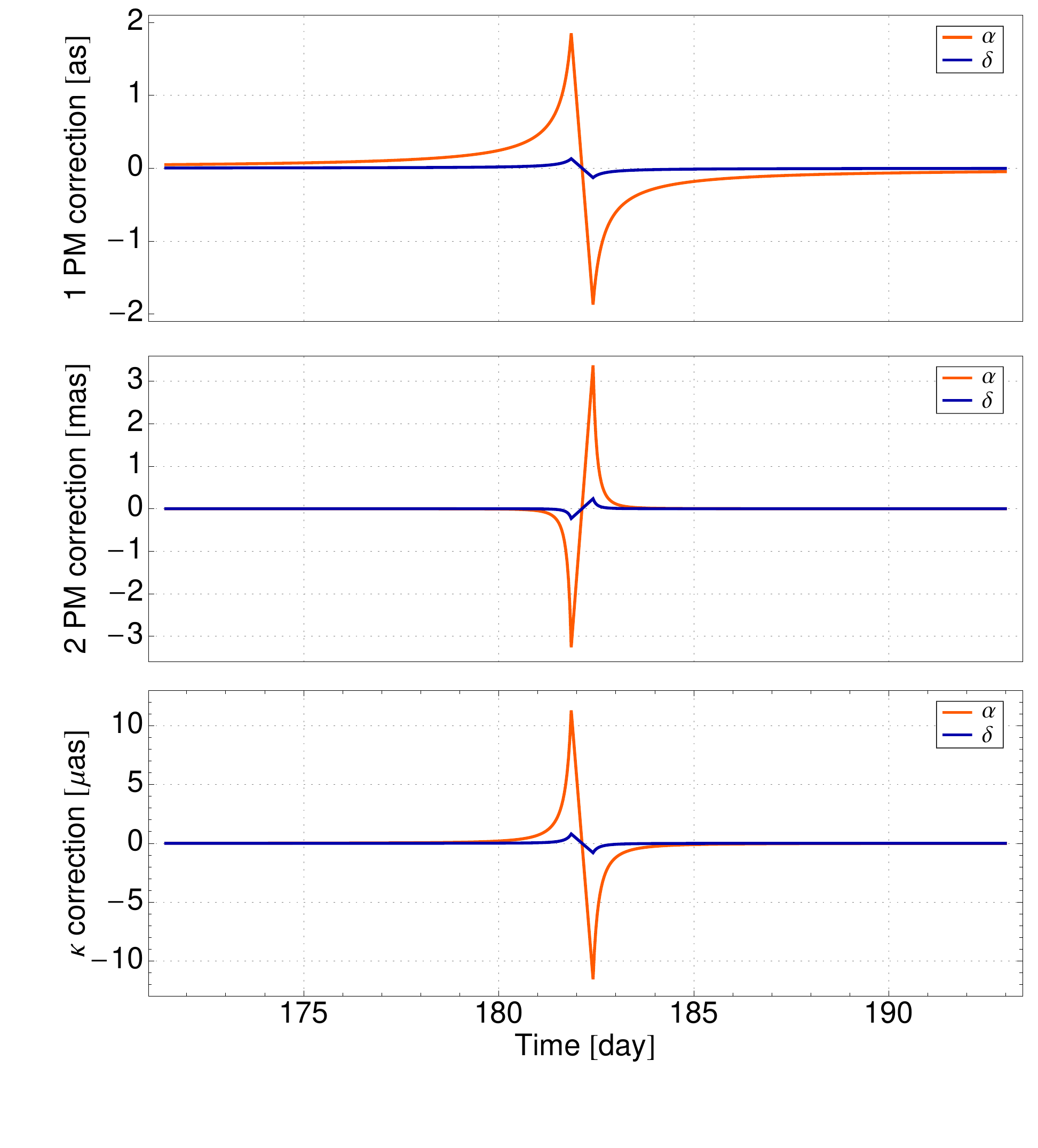}\hfill
\includegraphics[width=0.45\textwidth]{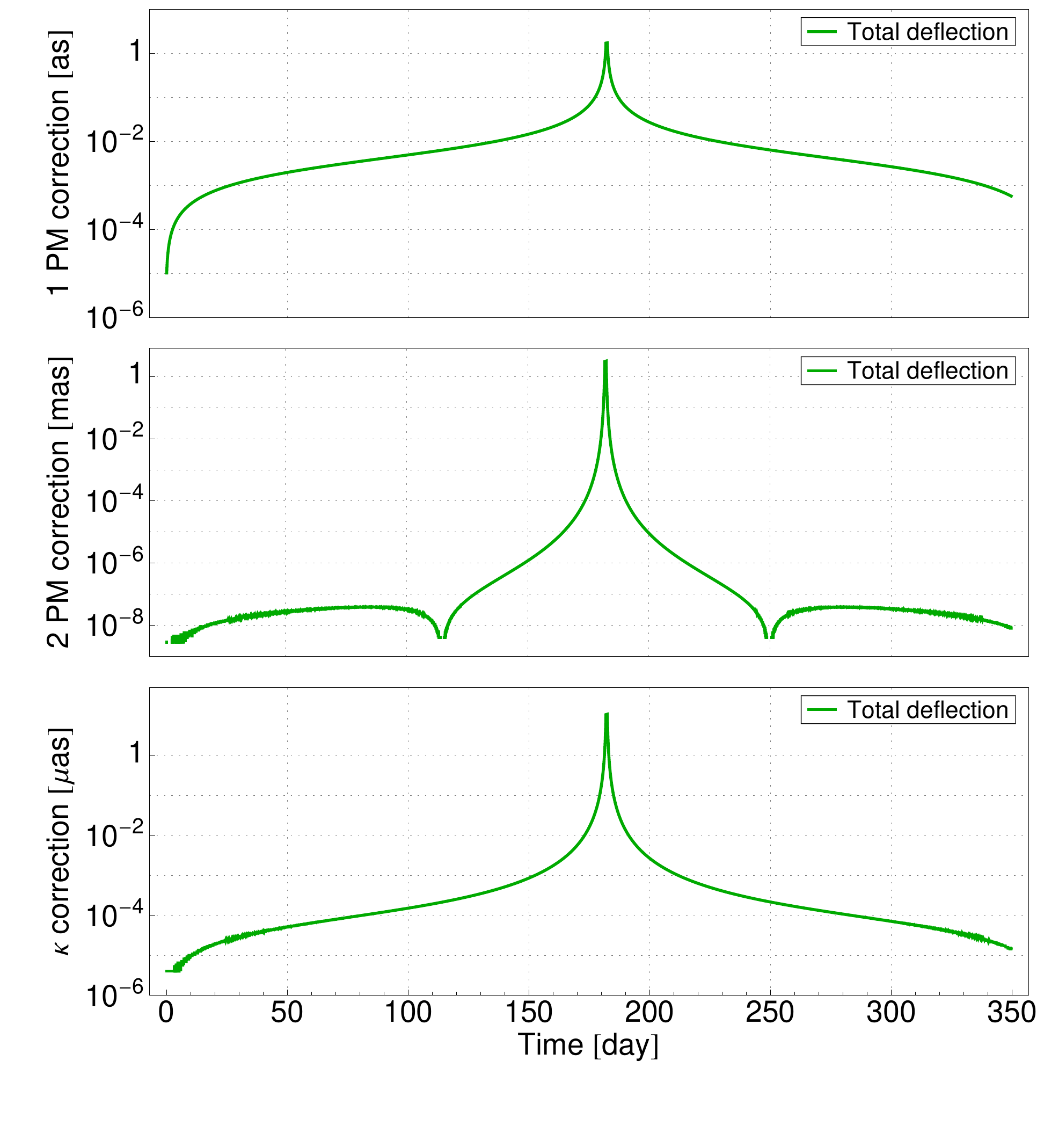}
\end{center}
\caption{Contributions to the observed direction of an incident light ray coming from a star. Left: contributions expressed for the right ascension and declination in the tetrad (see relation (\ref{eq:angles})) - Right: contribution to the total angular deflection. The 2PM contribution is the total formal 2PM contribution (included the so-called "enhanced 2PN terms"). The $\kappa$ contribution represents the $\kappa$ term in (\ref{eq:dDdxb2}).}
\label{fig:Astro}
\end{figure}

\vspace{1cm}
We then consider a hypothetical star located far away from the Solar System and nearly in the Earth's orbital plane. We compute the incident direction of the light ray emitted by this star and observed on Earth. The reference frame used to give the incident direction is given by a comoving kinematically nonrotating tetrad. The only gravitational interaction considered is the one of the Sun described by the metric (\ref{eq:schwarz}). The incident direction of the light ray can be computed using Eq.~(\ref{eq:dDdxb1}) and Eq.~(\ref{eq:dDdxb2}) into Eq.~(\ref{eq:astro}), and the expression of the tetrad (\ref{eq:tetrad}). The incident direction of the light ray with respect to the tetrad is denoted by $n^{(i)}$ and can be parametrized by two angles $\alpha$ and $\delta$ usually called right ascension and declination
\begin{equation}\label{eq:angles}
	n^{\left< i \right>}=(\cos \alpha \cos \delta,\sin \alpha \cos \delta ,\sin \delta).
\end{equation}
Figure~\ref{fig:Astro} represents the 1PM and 2PM contributions to $\alpha$ and $\delta$ as well as the total deflection angle. As one can see from relation~(\ref{eq:dDdxb2}), the 2PM correction to the angular measurement depends on two terms: a first term proportional to $\kappa$ and a second one proportional to $(1+\gamma)^2$, both of them being formally of order 2PM. Nevertheless, it is known that the term proportional to $(1+\gamma)^2$ can be absorbed in the 1PM term by a change of variable and it is therefore usually called "enhanced 2PN term" (for further details, see~\cite{klioner:2010fk,teyssandier:2012uq}). The enhanced 2PN term has a contribution of the order of few milliarcseconds (\rm{mas})  while the second order contribution proportional to $\kappa$ has a contribution of 10 microarcseconds ($\rm{\mu as}$) only.

\subsection{Angular distance between two stars as measured from Earth}
For this application, we consider two hypothetical stars located far away from the Solar System nearly in the Earth's orbital plane and we compute the angular separation between these two stars as measured from Earth. This representation can be used as a very simplified model of the GAME space mission~\cite{vecchiato:2009vn,gai:2012fk,gai:2012kx,gai:2012uq}. The only gravitational interaction considered here is the one due to the Sun. Relation (\ref{eq:ang_sep}), giving the angular separation between two incident light rays, can be simplified in the case of static and spherically geometry described by the space-time metric (\ref{eq:petric1}). The observed angle $\phi$ between two stars can then be written as
\begin{equation} \label{eq:angsep}
	\sin^2 \frac{\phi}{2}=\frac{1}{4}\left[\frac{\left(A(r_B)-B(r_B)\beta^2\right) |\hat{\bk}'-\hat{\bk}|^2}{B(r_B)(1+\beta^m\hat k_m)(1+\beta^l\hat k_l')}\right]_B \, ,
\end{equation}
where $\left(\hat{k}_j\right)_B$ and $\left(\hat{k}_j'\right)_B$ are the components of the deflection functions of the two incident light rays expressed in global coordinates that can be computed using the expression (\ref{2d1}), $A(r)$ and $B(r)$ are the functions parametrizing the metric (\ref{eq:petric1}) and $\beta^i=v^i/c$ is the coordinate velocity of the observer. We apply the last expression in a Schwarzschild geometry. The functions $A(r)$ and $B(r)$ are then given by (\ref{eq:schwarz}) and the $\hat \bk$ vectors are determined by (\ref{2d1}), once Eq.~(\ref{eq:dDdxb1}) and Eq.~(\ref{eq:dDdxb2}) have been introduced.
\begin{figure}[hbt]
\begin{center}
\includegraphics[width=\textwidth]{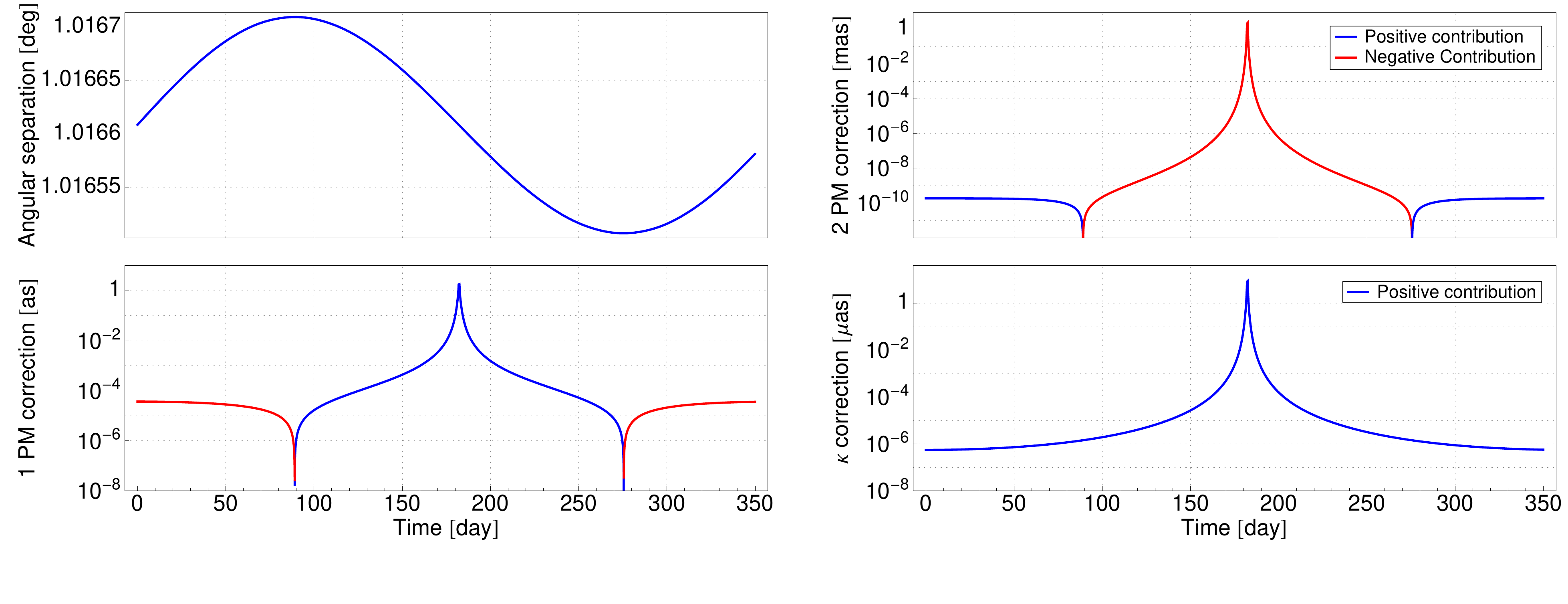}
\end{center}
\caption{Contributions to the angular separation between two incident light rays coming from two stars as observed from Earth. The 2PM contribution is the total formal 2PM contribution (included the so-called "enhanced 2PN terms"). The $\kappa$ contribution is the contribution proportional to the $\kappa$ term in (\ref{eq:dDdxb2}).}
\label{fig:Angular}
\end{figure}
Figure~\ref{fig:Angular} represents the evolution of the angular separation~\eqref{eq:angsep} with respect to time and the contribution of the 1PM and 2PM corrections. As for the direction of the incident light ray (see previous section), the 2PM correction to the angular measurement depends on two terms: a first term proportional to $\kappa$ and a second one proportional to $(1+\gamma)^2$. In this case too, the so called "enhanced 2PN" term has a contribution of the order of few \rm{mas} while the 2PM contribution, proportional to $\kappa$, has a contribution of 10 $\rm{\mu as}$ only.

We shall recall that the accuracy aimed by modern astrometric missions is about the $\mu as$ level, so that most 2PM order effects are observable near the Sun while the "enhanced 2PN term" also need to be taken into account when observing near Jupiter or Saturn.

\section{Conclusions}\label{sec:conclusions}
In this paper, we use the Time Transfer Function in order to compute range, Doppler and two kind of astrometric observables: the absolute incident direction of light rays in a given frame and the angular separation between two incident light rays. The formulation presented in Section~\ref{sec:dop_astr} is very general and can be used at any order. All the observables depend on the TTF and its derivatives. We also show how to numerically compute the TTF and its derivatives up to 2PM order. This is done in the form of integrals of functions of the metric and its derivatives taken along a straight line. This method is particularly efficient from a numerical point of view. On one hand, it does not require one to numerically derive the TTF (which can lead to numerical error). On the other hand, it does not require the computation of the full trajectory of the photon in curved space-time, which is a Boundary Value Problem (see \cite{san-miguel:2007hc}). This approach can be applied to any metric and therefore can also be used to determine observables in alternative theories of gravity (as long as the light propagation is governed by a null geodesic).
We also present a version of the formalism valid in the case of a static, spherically symmetric space-time. As a validation of our method, we explicitly compute analytically the TTF and its derivatives in the case of a Schwarzschild-like geometry and compare our expressions with well established results from~\citep{teyssandier:2012uq}. 
Finally, we apply our formulae to compute the Range and Doppler for a BepiColombo-like space mission and to simulate different configurations of a Gaia-like and GAME-like astrometric observations.
We show that the standard model used for radioscience measurements is accurate at a level just below BepiColombo accuracy. We also highlight that modern $\mu as$-astrometry needs to take into account second order relativistic corrections for observations near the limb of the Sun and of giant Solar System planets. 

\begin{acknowledgments}
The authors warmly thank P. Teyssandier for useful comments that helped to improve the readability of the manuscript. The research described in this paper was partially carried out at the Jet Propulsion Laboratory, California Institute of Technology, under contract with the National Aeronautics and Space Administration\ \copyright \ 2013. A.H. acknowledges support from the Belgian American Educational Foundation (BAEF) and from the Gustave-Bo\"el - Sofina "Plateforme pour l'Education et le Talent".
\noindent S.B. thanks the French-Italian University (UIF/UFI) for the financial support of this work.
\noindent S.B. and C.L.P.-L. are grateful for the financial support of CNRS/GRAM and CNES/Gaia.
\end{acknowledgments}

\bibliography{biblio_TTF}
\end{document}